\documentclass[12pt]{article}

\usepackage{scicite}
\usepackage{times}
\usepackage{amsfonts}
\usepackage{amsmath}
\usepackage{xcolor}
\usepackage{graphicx}
\newtheorem{lemma}{Lemma}
\newtheorem{theorem}{Theorem}
\usepackage[labelfont=bf,labelsep=period,figurename={Fig.},font=small]{caption}

\newenvironment{sciabstract}{%
	\begin{quote} \bf}
	{\end{quote}}

\title{A new class of higher-order topological insulators that localize energy at arbitrary multiple sites}

\author
{Yimeng Sun,$^{1}$ Linjuan Wang,$^{2\ast}$ Huiling Duan,$^{1}$ Jianxiang Wang$^{1\ast}$\\
	\\
	\normalsize{$^{1}$State Key Laboratory for Turbulence and Complex Systems,}\\
	\normalsize{Department of Mechanics and Engineering Science,}\\
	\normalsize{College of Engineering, Peking University, Beijing 100871, China}\\
	\normalsize{$^{2}$School of Astronautics, Beihang University, Beijing 100191, China}\\
	\\
	\normalsize{$^\ast$ Corresponding author. Email: wanglj@buaa.edu.cn; jxwang@pku.edu.cn}
}

\date{}

\begin{document}
	
	
	
	
	\maketitle

	\begin{sciabstract}
		$\mathbb{Z}$-classified topological phases lead to larger-than-unity topological states. 
		However, these multiple topological states are only localized at the corners in nonlocal systems. Here, first, we rigorously
		prove that the multiple topological states of nonlocal Su-Schrieffer-Heeger (SSH) chains can be inherited and realized by local aperiodic chains with only the nearest couplings. Then, we report a new class of higher-order topological insulators constructed with the local aperiodic chains, which can have any integer number of 0D topological states localized at arbitrary positions in the whole domain of the insulators, including within the bulk.
		The 0D topological states are protected by the local topological marker in each direction, instead of the bulk multipole chiral numbers in the existing work.
		Our work provides multiple combinations of localized corner-bulk topological states, which enables programmable lasers and sasers by selecting the excitation sites without altering the structure, and thus opens a new avenue to signal enhancement for computing and sensing.
	\end{sciabstract}
	
	
	\section*{Main Text}
	\subsection*{Introduction}
	Topological phases lead to robust topological states and have lasting impacts in the fields of electronics \cite{Hasan2010,ZhangSC2011,semimetal2018,Schnyder2008,Chiu2016,sanjuti-jixian}, photonics \cite{photonic2008,photonic2013,Photonic2021,Wang2023,newcornerstate,kirsch_nonlinear_2021,pan_real_2023} and phononics \cite{Ma2019,Xue2022,Xiao2015exp,Wang_2022,local-marker,highorder,PRLpumping,nj-zs,pnas-mecha,pumpelastic}.  $\mathbb{Z}$-classified topological phases, exotic phenomena, may endow topological systems with any integer number of topological states at interfaces, which is more flexible than one topological state in terms of broadband \cite{Jiang22,multi} and multi-position energy localization, and thus have been pursued~\cite{multipolar_lasing,duotiao,degenerate-disclination,Zhang2023}. To date, 
	0D topological states in both higher-order $\mathbb{Z}_2$-classified and $\mathbb{Z}$-classified topological systems
	are only realized at corners of the systems, though the nonlocal interactions in the static topological systems~\cite{Z-the,Z-exp} and time-periodic modulations in the Floquet topological systems~\cite{Floquetsum}
	enable more topological states. Compared to nonlocal systems \cite{any-roton,indexing,Haldane2014}, local ones can avoid coupling inversion \cite{Z-exp}, complexity of the geometric structure, and geometric overlapping of dimensions caused by implementing nonlocal interactions \cite{chen_roton-like_2021,exp-roton-like,CM-nonlocal-2022}. 
	However, multiple 0D topological states induced by $\mathbb{Z}$-classified topological phases with large winding numbers have not been realized in static systems with local interactions, let alone in the bulk. Real robust bulk states (not the ones at the interfaces of two media) are being pursued, since they can be customized and can make full use of the footprint \cite{Wang_2022,CABS-prl}.

	In this work, we for the first time report a new class of $\mathbb{Z}$-classified higher-order topological insulators
	constructed with the 1D local aperiodic chains. This class of $\mathbb{Z}$-classified topological phases is characterized by the local topological marker in each direction, instead of the bulk multipole chiral numbers in the existing work~\cite{Z-the,Z-exp}. In contrast to previous findings that 0D topological states are only localized at corners, we demonstrate that
	0D topological states can be realized at arbitrary multiple positions in the whole domain of the  topological insulators.
	We experimentally observe the multiple 0D topological states  in a 2D local acoustic practical structure. These multiple 0D topological states can customize energy localization at  arbitrary multi-positions, which has applications in signal enhancement for computing and sensing.
	
	\subsection*{Results}
	\subsubsection*{Multiple topological states in 1D systems and the realization}
\begin{figure}
		\centering
		\includegraphics[width=0.8\linewidth]{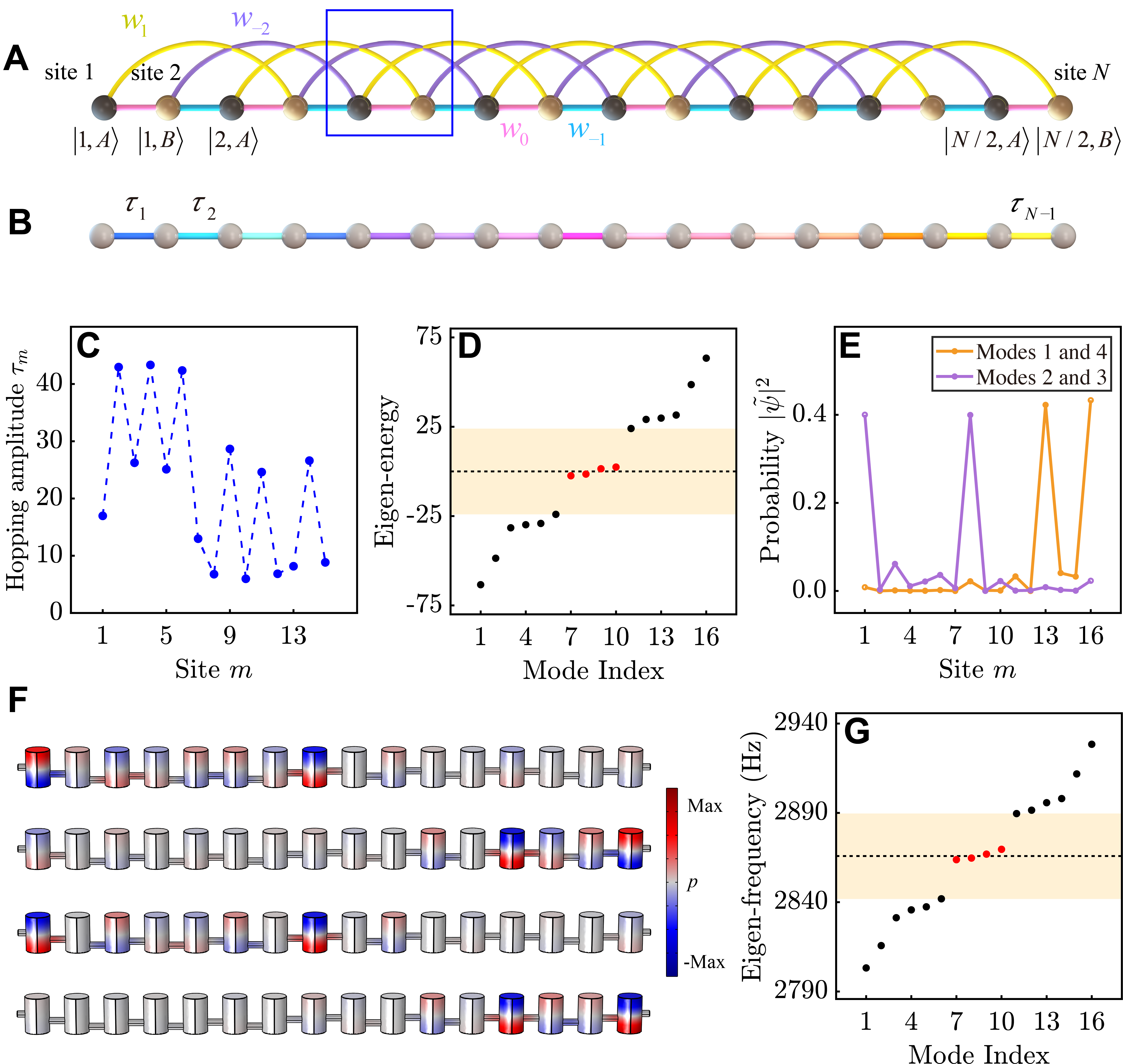}
		\caption{\label{Fig-1} \textbf{The multiple topological states in 1D systems.} (\textbf{A}) Schematic diagram of an electron extended SSH chain with long-range hoppings shown as ``bonds'' $w_n\ (n=-n_\mathrm{left}, \cdots, n_\mathrm{right})$. The chain consists of eight unit cells (one in the frame of blue solid line), with two sublattice sites of type $A$ [black dots, i.e., sites $(l,A)$] and type $B$ [brown dots, i.e., sites $(l,B)$]. Also, we use the ``$m$th site'' to denote site $(l,A)$ with $m=2l-1$ or site $(l,B)$ with $m=2l$. (\textbf{B}) Schematic diagram of the aperiodic chain with the nearest-neighbor hopping amplitudes $\tau_1, \tau_2, \cdots, \tau_n$. (\textbf{C}) The hopping amplitudes of the aperiodic chain, which is mapped from the extended SSH chain ($w_{-2}=33, w_{-1}=w_0=w_1=12$). (\textbf{D}) The energy spectra of the extended SSH chain and also of the aperiodic chain, which are overlapped, where red dots correspond to the topological states. (\textbf{E}) Four topological zero modes of the aperiodic chain. Two of them are localized at the first and eighth sites; and the other two at the thirteenth and sixteenth sites. (\textbf{F}) The designed acoustic coupled-resonators system with sixteen coupled resonators  [each row in (F)] numbered one to sixteen from left to right, and four topological ``zero modes'' obtained by finite element software COMSOL Multiphysics. (\textbf{G}) The frequency spectrum of the system, where red dots correspond to the topological states.}
	\end{figure}

	We consider a 1D nonlocal chain -- noninteracting, spinless electron extended SSH chain \cite{CHEN2020,Electronlong2019} with bipartite long-range hoppings (hopping between sublattice $A$ and sublattice $B$), shown in Fig.~\ref{Fig-1}A. 
The picture shows an example only with the nearest-neighbor hopping, and the third-nearest-neighbor long-range hopping, but our theoretical analysis to follow is for any bipartite long-range hoppings.
	The bulk real-space Hamiltonian of the chain with $N$ sites ($N/2$ unit cells) under the open boundary conditions is
	\begin{equation}
		\label{eq1}
		\hat{H}=\sum_{l=1}^{N/2}\sum_{n=-n_\mathrm{left}}^{n_\mathrm{right}} w_n^*\vert l,A \rangle \langle l+n,B \vert +\sum_{l=1}^{N/2}\sum_{n=-n_\mathrm{left}}^{n_\mathrm{right}} w_n\vert l+n,B \rangle \langle l,A \vert,
	\end{equation}
	where $ 1\leq l+n \leq N/2$; $\vert l,a \rangle$, with $l\in \{1,2,\cdots,N/2\}$ and $a\in\{A,B\}$, denotes the state of the chain where the electron is in the $l$th unit cell on the site at sublattice $a$; $w_n^{*}$ denotes the hopping amplitude from $B$ in the $(l+n)$th cell to $A$ in the $l$th cell; and $w_n$ denotes the hopping amplitude from $A$ in the $l$th cell to $B$ in the $(l+n)$th cell. Here we take the hopping amplitudes to be real and non-negative. The eigenequation is
	\begin{equation}
		\label{eq2}
		\hat{H}\vert \psi \rangle=E\vert \psi \rangle,
	\end{equation}
	where $\vert \psi \rangle$ denotes the wave function, and $E$ represents the eigen-energy. For convenience of derivation and calculation, all operators for the 1D chain (e.g.,~$\hat{H}$) are written in the same position representation (the basis is in the order of $\vert 1,A\rangle, \vert 1,B\rangle, \cdots, \vert N/2,A\rangle, \vert N/2,B\rangle$). Since the bipartite electron extended SSH chain possessing chiral symmetry $H\Gamma=-\Gamma H$ \cite{Asbth2016,Maffei2018}, where $\Gamma=I_{(N/2)\times(N/2)}\otimes \sigma_z$ ($\sigma_z$ is the Pauli matrix and $\otimes$ is the Kronecker product), belongs to the BD${\rm \uppercase\expandafter{\romannumeral1}}$ class,  the spectrum of the chain is symmetric about the chiral symmetry axis $E=0$, and the number of zero-energy edge states which can only localize at the left or right edge should correspond to the value of the bulk winding number (defined in the vector space, see Supplementary Text \ref{a}) in the thermodynamic limit \cite{CHEN2020}. In addition, in the real space, the topological properties of the finite chain can be characterized by the local topological marker (see Supplementary Text \ref{a}), whose value equals the bulk winding number away from the edges.
	
	We map the above bipartite extended SSH chain onto a local aperiodic chain shown in Fig.~\ref{Fig-1}B using improved Householder tridiagonalization. The Householder method can transform a Hermitian (including real symmetric) matrix into a tridiagonal real symmetric matrix, which can be further transformed by elementary row and column transformations so that the subdiagonal elements are positive, while the main-diagonal elements remain unchanged (the tridiagonalization begins from the first row and column of the symmetric matrix; process is in Supplementary Text \ref{b}). Compared to the Lanczos tridiagonalization \cite{lanczosnature,pnas2021}, the Householder method has numerical stability when the dimension of the matrix increases. Here, we find an important property of the Householder method, namely, {\it for a Hermitian matrix $H_{n\times n}$, the tridiagonalized matrix by the Householder method remains the same, when simultaneously exchanging any two rows and the corresponding two columns, except for the first row and first column, of $H$} (Lemma~\ref{Theorem-1} in Supplementary Text \ref{c}),
	considering the nontrivial case that the tridiagonalized matrix has nonzero subdiagonal elements.
	
	By using the Householder tridiagonalization, when the extended SSH chain with Hamiltonian $H$ is mapped onto the local aperiodic chain with tridiagonal Hamiltonian $H_{\mathrm{eff}}=V^{\dagger}HV$, where $V$ is a unitary matrix satisfying $V^{\dagger}V=I$, the eigenvalues of the original Hamiltonian and the orthogonality of the eigenvectors do not change. The relation between the eigenvector $\tilde{\psi}$ after transformation and the eigenvector $\psi$ before transformation is $\tilde{\psi}=V^{\dagger}\psi$.  In the traditional Householder method \cite{householder,recipe}, $V_{11}=1$; thus the dynamics of the first site in the original extended SSH chain is faithfully mapped to the first site in the aperiodic chain; in this sense,
	the first site in the original chain is called ``the anchor site'' \cite{lanczosnature,pnas2021}, with the dynamics completely preserved. Here, in order to examine the dynamics of any site
	in the original chain that has multiple topological states, we improve the traditional Householder method to enable any site to become ``the anchor site''.
	In virtue of Lemma~\ref{Theorem-1}, we prove the following theorem (proof is given in Supplementary Text \ref{c}):
	\begin{theorem}
		\label{Theorem-2}
		All on-site potentials of the tridiagonalized chain unitarily mapped from any chiral-symmetric system are zero for any chosen anchor site.
	\end{theorem}Here, a chiral-symmetric system means that there exists a position basis in which its Hamiltonian matrix satisfies $\Gamma H=-H\Gamma$. For a chiral-symmetric system represented by the schematic diagram such as Fig.~\ref{Fig-1}A, every closed circuit composed of hopping ``bonds'' contains an even number of sites. The chiral-symmetric system can be nonlocal or high-dimensional.
	
	As an example, the improved Householder transformation is carried out on the finite chain that has eight unit cells with hopping amplitudes  $w_{-2}=33, w_{-1}=w_0=w_1=12$ and a bulk winding number 2, shown in Fig.~\ref{Fig-1}A. The nearest-neighbor hopping amplitudes of the mapped aperiodic chain are calculated as shown in Fig.~\ref{Fig-1}C, with the first site, i.e.,~($1,A$) being the anchor site. The energy spectrum of the aperiodic chain is shown in Fig.~\ref{Fig-1}D, where there are four topological states at zero energy protected by special nonlocal symmetry (see discussions below), and they are slightly nondegenerate due to the finite number of unit cells. The four topological states after transformation are shown in Fig.~\ref{Fig-1}E: a pair of edge states originally localized at the first and sixteenth sites in the original extended SSH chain is mapped to new locations at the first and eighth sites after transformation, and a pair of edge states originally localized at the third and fourteenth sites is localized at the thirteenth and sixteenth sites after transformation. Each pair of topological states after transformation satisfies $\tilde{\psi}_2=\Gamma\tilde{\psi}_1$, so $\tilde{\psi}_2$ and $\tilde{\psi}_1$ are localized at the same sites but have different phase distributions (explanation is given in Supplementary Text \ref{c}). The definition of localization for topological states in aperiodic chains is different from that in periodic systems where the states decay exponentially into the bulk as a function of distance; instead, we use the inverse participation ratio to quantify the localization degree of the states in aperiodic chains. The localization degree of each pair of topological states in the mapped aperiodic chain is close to that in the original chain (compared in Supplementary Text \ref{d}).  It is worth noting that after the Householder transformation, except for the first site whose dynamics is faithfully mapped, the localization degrees of the other three sites (i.e., sites 8, 13 and 16) are determined by the selection of matrix $V$. Selecting a suitable anchor site can result in high localization degree, so the anchor site is generally selected as one of the localized sites in the original chain. It is noted that aperiodic topological chains with artificial dimensions typically have about 20 sites \cite{pnas2021,lanczosnature}. Our improved Householder method is accurate for relatively large chains with more than 100 sites, and it is also numerically stable, which is detailed in Supplementary Text \ref{z4}.
	
	Now, we examine the topological markers of the mapped local aperiodic chain. The topological marker is totally determined by the tridiagonal Hamiltonian $H_{\mathrm{eff}}$ and the unitary matrix $V$.
	The topological states of the aperiodic chain are protected by a special ``nonlocal symmetry'', $\tilde{\Gamma}=V^{\dagger}\Gamma V$, so that the tridiagonal Hamiltonian satisfies $\tilde{\Gamma}H_{\mathrm{eff}}=-H_{\mathrm{eff}}\tilde{\Gamma}$. However, as $V$ depends on and varies with the Hamiltonian of the original chain, $H$, here $\tilde{\Gamma}$ is not a constant operator as in a usual symmetry relation; rather, the determination of $\tilde{\Gamma}$, or the specific form of such ``nonlocal symmetry'' for $H_{\mathrm{eff}}$, is ultimately based on the original $H$, and the original symmetry relation $\Gamma H = -H \Gamma$. The projectors $\Gamma_{A}$ and $\Gamma_{B}$ on sublattices $A$ and $B$ are transformed into $\tilde{\Gamma}_{A}=V^{\dagger}\Gamma_{A} V$ and $\tilde{\Gamma}_{B}=V^{\dagger}\Gamma_{B} V$, respectively. Under the special nonlocal symmetry, the transformed local topological maker for the aperiodic chain is defined in the real space as follows. All the eigenvectors of $H_{\mathrm{eff}}$ are arranged according to the corresponding eigenvalues in an ascending order to obtain $\tilde{\psi}_{-}=[\tilde{\psi}_1,\tilde{\psi}_2,\cdots,\tilde{\psi}_{N/2}]$ and $\tilde{\psi}_{+}=[\tilde{\psi}_{N/2+1},\tilde{\psi}_{N/2+2},\cdots,\tilde{\psi}_{N}]$. Then, the ``flat-band Hamiltonian'' $\tilde{Q}$ is defined as $\tilde{Q}=\tilde{P}_+-\tilde{P}_-$, where $\tilde{P}_{+}=\tilde{\psi}_{+}\tilde{\psi}_{+}^{\dagger}$ and $\tilde{P}_{-}=\tilde{\psi}_{-}\tilde{\psi}_{-}^{\dagger}$. Having $\tilde{Q}$ and $\tilde{\Gamma}_{A(B)}$, $\tilde{Q}_{AB}=\tilde{\Gamma}_{A} \tilde{Q} \tilde{\Gamma}_{B}$ and $\tilde{Q}_{BA}=\tilde{\Gamma}_{B} \tilde{Q} \tilde{\Gamma}_{A}$ can be given. Finally, the transformed local topological marker is expressed as
	\begin{equation}
		\label{eq3}
		\tilde{\gamma}(l)=\frac{1}{2} \sum\limits_{a=A,B} V\{(\tilde{Q}_{BA}[\tilde{X},\tilde{Q}_{AB}])_{la,la}+ (\tilde{Q}_{AB}[\tilde{Q}_{BA},\tilde{X}])_{la,la} \}V^{\dagger},
	\end{equation}
	where $\tilde{X}$  is defined as $\tilde{X}=V^{\dagger}XV$ ($X$ is the position operator of the original extended SSH chain), and $la$ represents the position of a site. From Eq.~\ref{eq3}, we prove that the topological marker of the aperiodic chain has the same value as that of the original extended SSH chain (proof is given in Supplementary Text \ref{e}), and the topological marker away from the edges of the aperiodic chain can determine the topological property of the chain, as can the original extended SSH chain. Therefore, the class of local aperiodic chains inherit the topological properties of the chiral-symmetric extended SSH chains. We note that the topological properties of the new local chain are clearly determined as long as the matrix $V$ defining the symmetry transformation is given; however, it is a question of interest whether there exists a unitary transformation (if $V$ not known a priori) that could associate a given tridiagonal Hamiltonian $H_\mathrm{eff}$ with a Hamiltonian of some periodic lattice structure. Although this is a challenging mathematical question, achieving the inverse transformation when $V$ is unknown is not essential for realizing multimode topological bulk states in a local system, which is thus not a concern of the studies of topological physics \cite{lanczosnature,pnas2021,highorder,zhu_time-periodic_2022}.
	
	Based on the above theoretical framework, we design a simple 1D practical structure -- 1D acoustic coupled-resonators system according to the local aperiodic chain to demonstrate the existence of its multiple topological states. In terms of the mapped aperiodic chain in Fig.~\ref{Fig-1}B, sixteen acoustic coupled-resonators with the same height of 60 mm, and with the same on-site frequency to simulate the same on-site potential, are equally placed and connected by tubes to form a chain, shown in Fig.~\ref{Fig-1}F, and the couplings (hoppings) are introduced by changing the vertical positions of the connecting tubes between every two nearest resonators (Fig.~\ref{Fig-1}F). The sizes of the resonators and the connecting tubes are listed in Materials and Methods.
	The hopping amplitudes between every two neighboring resonators are determined by the vertical positions of the connecting tubes.
	For the given values of the hopping amplitudes in Fig.~\ref{Fig-1}\textit{C}, we determine the corresponding heights of the connecting tubes by the finite-element software COMSOL Multiphysics. The corresponding heights of the connecting tubes from left to right in the system are shown in Materials and Methods. The spectrum corresponding to the sixteen eigenmodes is shown in Fig.~\ref{Fig-1}G, where the four red points in the center correspond to the topological states whose modes are shown in Fig.~\ref{Fig-1}F. We see that Fig.~\ref{Fig-1}G well emulates the spectrum in Fig.~\ref{Fig-1}D. From Fig.~\ref{Fig-1}F, the four topological states are divided into two pairs, one of which is localized at the first and eighth sites, but the phase distributions are different concerning these two sites, exhibiting symmetric (third row in Fig.~\ref{Fig-1}F) and anti-symmetric (first row) distributions; the other pair is localized at the thirteenth and sixteenth sites, and the phase distributions are different.
	
\begin{figure}
		\centering
		\includegraphics[width=0.75\linewidth]{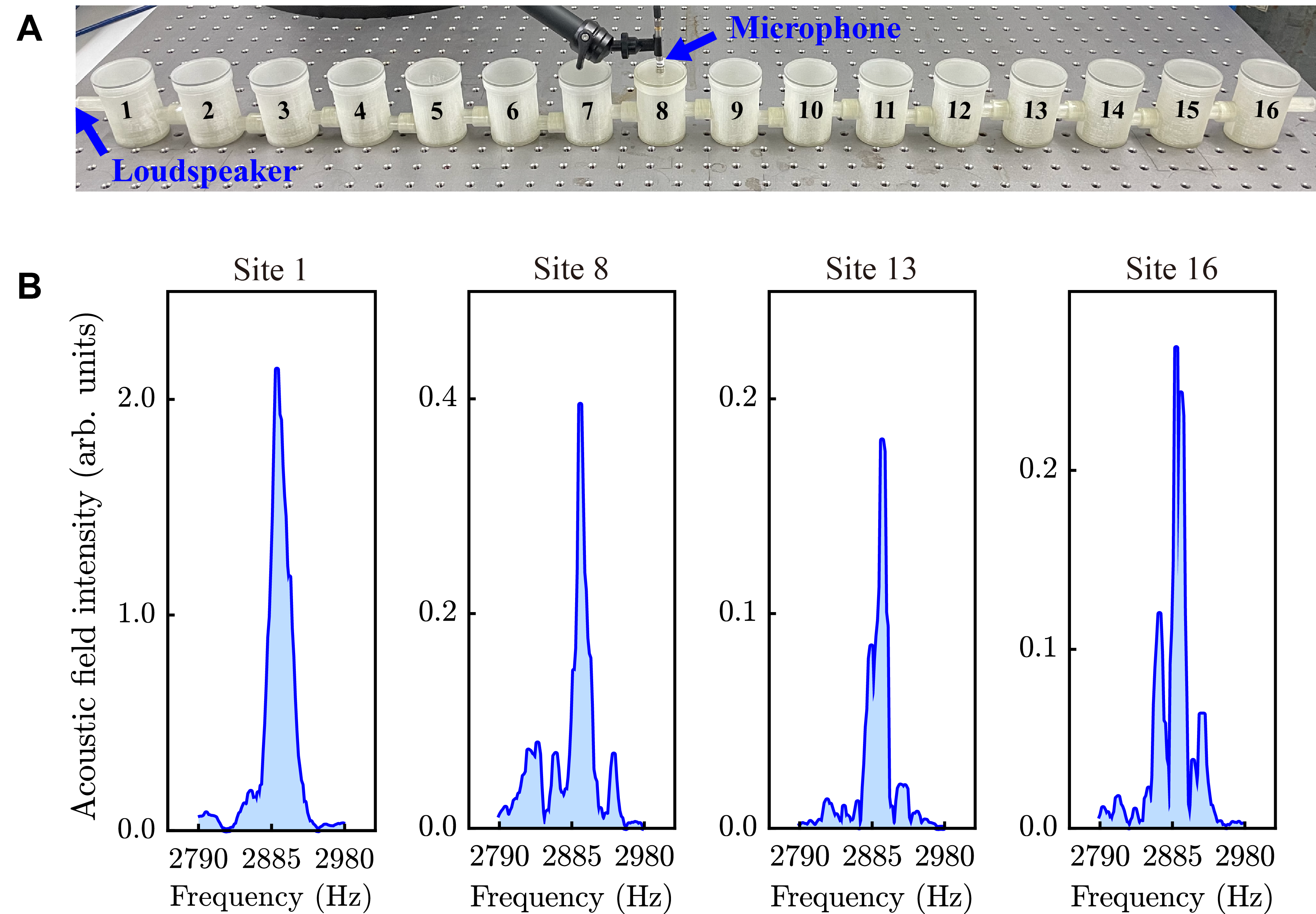}
		\caption{\label{Fig-2} \textbf{Fabricated acoustic coupled-resonators system and measured acoustic intensity response spectra.} (\textbf{A}) Photograph of the experimental system. For the measurement of intensity response spectra, the sound source is input from the port on the leftmost side of the whole structure, and the microphone measures the sound pressure through small ports on the upper surfaces of the resonators at the first, eighth, thirteenth and sixteenth sites [only the measurement for the eighth site is shown in (A)]. (\textbf{B}) The acoustic intensity response spectra. Sharp transmission peaks are observed at frequency range 2890 Hz - 2900 Hz for the four sites.}
	\end{figure}

\begin{figure}
		\centering
		\includegraphics[width=0.75\linewidth]{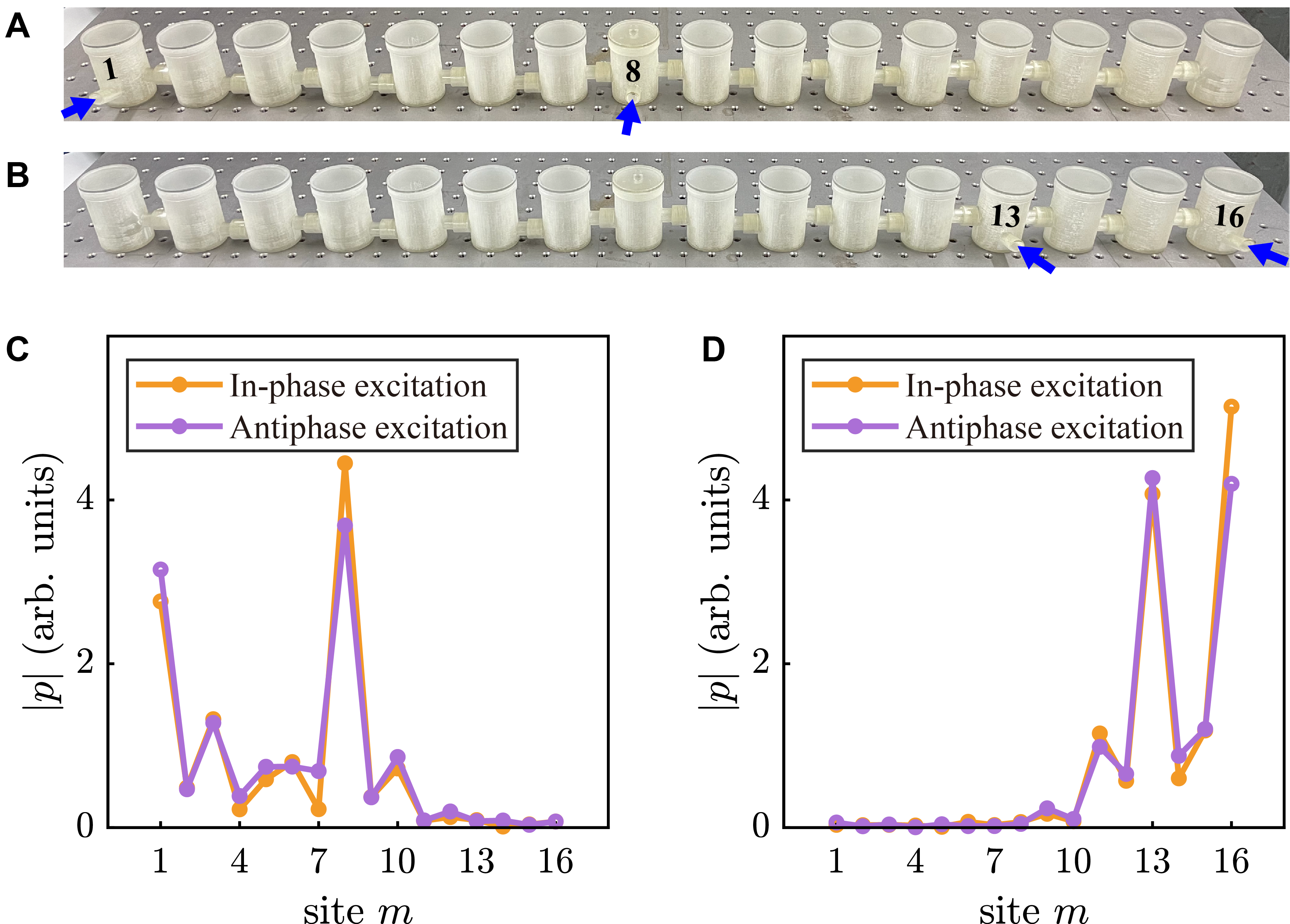}
		\caption{\label{Fig-3} \textbf{Experimental observation of the spatial sound field distribution for topological states.} (\textbf{A}) Sample for adding selected (in-phase or anti-phase) excitation sources at the first and eighth sites. The blue arrows represent the inputs of sound sources through the front ports. (\textbf{B}) Sample for selected excitation sources at the thirteenth and sixteenth sites. (\textbf{C}) Two topological states localized at the first and eighth sites achieved by selected excitations. The state excited by in-phase sources at these two sites is represented in orange, and that by anti-phase sources is in purple. (\textbf{D}) Two topological states localized at the thirteenth and sixteenth sites achieved by selected excitations.}
	\end{figure}

	Finally, we  demonstrate the predicted and simulated multiple topological states experimentally by fabricating the above acoustic coupled-resonators system. The acoustic coupled-resonators system is fabricated by three-dimensional stereolithography technique printed parts which are interlocked by screw threads in sequence (experimental details are given in Materials and Methods), and the thicknesses of the materials are sufficient to ensure the boundary conditions of the hard sound field. First, a sound source is input from the left port of the structure (Fig.~\ref{Fig-2}A).
	In the frequency range 2790--2980 Hz (Fig.~\ref{Fig-2}B), a single-frequency source with a fixed amplitude is given at each time; the frequency is increased by 2 Hz each time from 2790 Hz. The sound pressure at each frequency is measured by a microphone at the first, eighth, thirteenth and sixteenth sites. The final sound intensity response spectra at these four sites are shown in Fig.~\ref{Fig-2}B. Figure \ref{Fig-2}B shows that sharp transmission peaks appear in the range 2890--2900 Hz (peak deviation of 30 Hz compared with the finite-element results 2860--2870 Hz, giving an error about 1$\%$) for the four sites, indicating that there are modes in the range 2890--2900 Hz.
	Second, we select suitable excitation sources to selectively excite the eigenmodes of the system. In Fig.~\ref{Fig-1}F, the simulated eigenmodes show that the system has two pairs of modes, where each pair is localized at the same sites but with different phase distributions. Therefore, in the experiment, excitation sources with the same phase (phase difference is 0) and opposite phases (phase difference is $\pi$) at the two localized sites should be given to verify whether the system has such a pair of modes. To add the sound sources, two structures in Fig.~\ref{Fig-3}A and B are fabricated, where the front ports are opened at the first and eighth (Fig.~\ref{Fig-3}A), and thirteenth and sixteenth resonators (Fig.~\ref{Fig-3}B) (the blue arrows in Fig.~\ref{Fig-3}A and B indicate the locations of the in-phase and anti-phase sound sources at the front ports).	
The microphone is used to measure the sound pressure at each site. As shown in Fig.~\ref{Fig-3}C and D, a pair of modes is localized at the first and eighth sites, and the other pair at the thirteenth and sixteenth sites, showing typical characteristics of topological states. These observations demonstrate that the two modes with different phase distributions in each pair of modes can be selectively excited. The experimental results are consistent with the simulated modes excited by in-phase (anti-phase) sources (in Supplementary Text \ref{h}), that is, the one-dimensional aperiodic chain obtained by tridiagonalization has four localized topological states.

	\subsubsection*{Multiple 0D topological states in 2D systems and the realization}
\begin{figure}
		\centering
		\includegraphics[width=\linewidth]{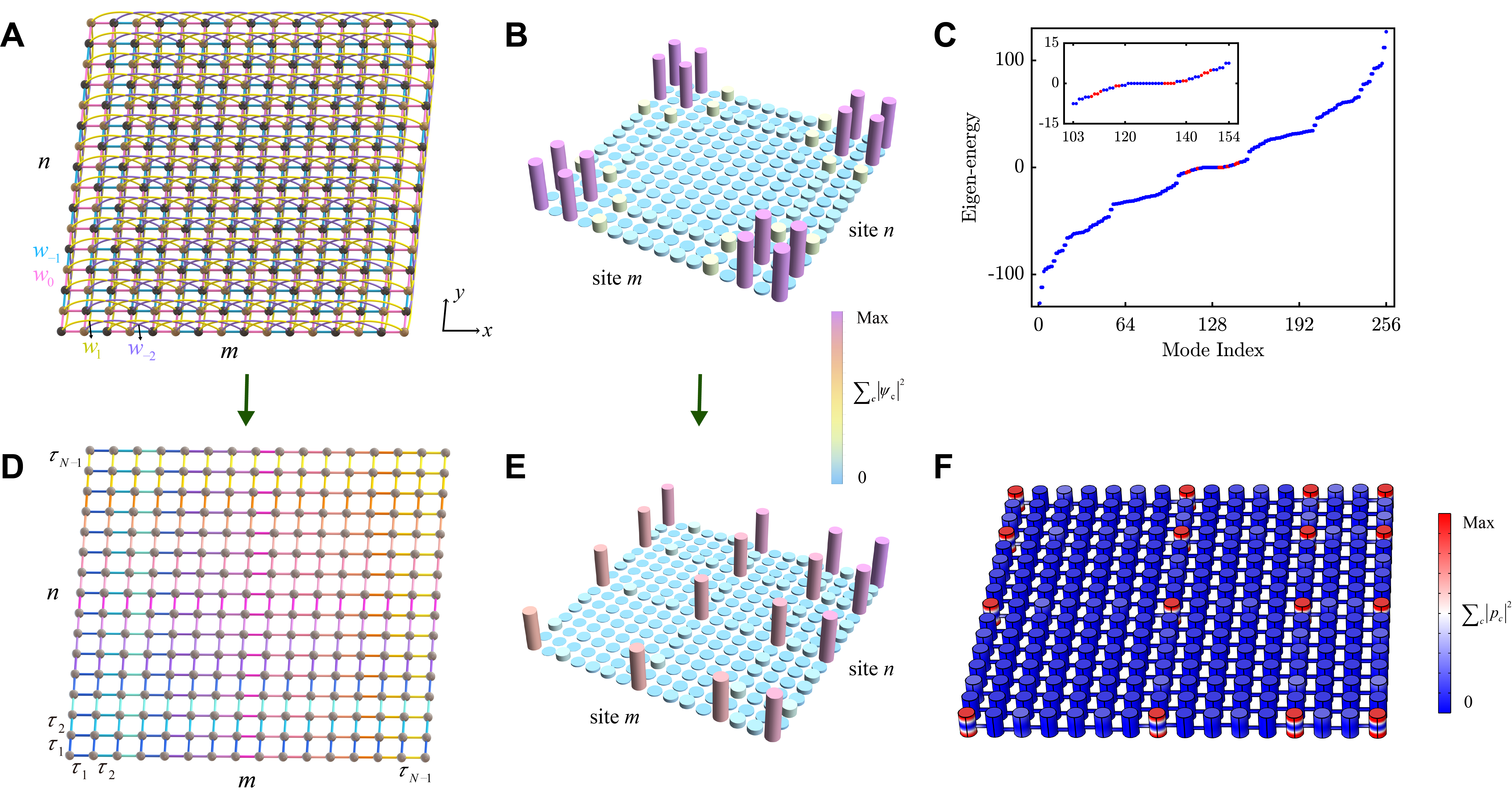}
		\caption{\label{Fig-4} \textbf{The multiple localized 0D topological states in 2D systems.} (\textbf{A}) Schematic diagram of the 2D extended SSH model with the long-range hoppings. The nearest-neighbor and third-nearest-neighbor hoppings are shown as an example. (\textbf{B}) The corner states of the 2D extended SSH model determined by winding number two are localized at sixteen sites. (B) shows the superposition of the probability distributions of the sixteen corner states. (\textbf{C}) The energy spectrum of the 2D extended SSH model, where the red dots represent the sixteen corner states. (\textbf{D}) Schematic diagram of the equivalent 2D local aperiodic model. (\textbf{E}) The 0D topological states of the 2D local aperiodic model are also localized at sixteen sites, but not only localized at the corners. (\textbf{F}) The 2D acoustic coupled-resonators system which is constructed according to (D). The superposition of sound intensities of sixteen 0D topological states has amplitudes at sixteen resonators, which is consistent with the result in (E).}
	\end{figure}

	Here, before we design higher-order local topological insulators which have localized 0D topological states in the bulk, we first construct a 2D nonlocal higher-order topological system -- the 2D extended SSH model as shown in Fig.~\ref{Fig-4}A by the bipartite 1D extended SSH chain in Fig.~\ref{Fig-1}A. We show that this model has multiple topological corner states per corner protected by $\mathbb{Z}$-classified
	topological phases which are characterized by the 1D winding number in each direction.
	
	Every row or column of the 2D extended SSH model  in Fig.~\ref{Fig-4}A is a 1D extended SSH chain in Fig.~\ref{Fig-1}A, so the Hamiltonian of the 2D extended SSH model is
	\begin{equation}
		\label{eq4}
		H^{\mathrm{2D}}=H\otimes I+I\otimes H,
	\end{equation}
	where $H$ is given in Eq.~\ref{eq1}. The 2D extended SSH model in Fig.~\ref{Fig-4}A has long-range hoppings between two sublattices of type $A$  (black dots) and type $B$ (brown dots), so it satisfies chiral symmetry $\Gamma^{\mathrm{2D}}H^{\mathrm{2D}}=-H^{\mathrm{2D}}\Gamma^{\mathrm{2D}}$, where $\Gamma^{\mathrm{2D}}=\Gamma \otimes \Gamma$. Also, the 2D extended SSH model satisfies $D_4$ symmetry  (the $D_4$ symmetry includes the inversion symmetry, so $L^{\mathrm{2D}}H^{\mathrm{2D}}= H^{\mathrm{2D}}L^{\mathrm{2D}}$, where $ L^{\mathrm{2D}}=L\otimes L$); and it satisfies the time-reversal symmetry, i.e., $T^{\mathrm{2D}}H^{\mathrm{2D}}= H^{\mathrm{2D}}T^{\mathrm{2D}}$, where $ T^{\mathrm{2D}}=\mathcal{K}$. When $L^{\mathrm{2D}}$- and $T^{\mathrm{2D}}$-symmetry are both present, the Berry curvature of the 2D extended SSH model is 0 \cite{Miert-zak-symmetry}, so the topological property cannot be characterized by the Chern number \cite{PRL2DSSH}; instead, the topological property of the 2D extended SSH model depends on the corresponding 1D extended SSH chain and the winding number $\gamma$. There exist $\gamma^2$ topological corner states per corner.
	
	Therefore, unlike the previous 2D quadrilateral nontrivial higher-order topological insulators with at most four corner states \cite{science-polarization, PRL2DSSH,PRBnnc-circuit, nnc-phasediagram}, the 2D extended SSH model in Fig.~\ref{Fig-4}A has sixteen corner states, when the parameters of the 2D extended SSH model are the same as those of the 1D extended SSH chain with winding number two in Fig.~\ref{Fig-1}, i.e., $w_{-2}=33, w_{-1}=w_0=w_1=12$. These corner states are localized at sixteen sites as shown in Fig.~\ref{Fig-4}B, and the localization sites are $(m, n) \in \{1,3,14,16\}\times \{1,3,14,16\}$. The energy spectrum of the model in Fig.~\ref{Fig-4}A is shown in Fig.~\ref{Fig-4}C, which is chiral-symmetric about zero energy, and the red dots correspond to the localized corner states. Because the sixteen corner states are buried in the bulk states, weak on-site potentials are added at the sixteen localization sites to distinguish the corner states. The reason for the existence of localized corner states is that if the eigenvalues and eigenmodes of the 1D extended SSH chain in Eq.~\ref{eq2} are $E_i\vert \psi_i \rangle$ and $ E_j\vert \psi_j \rangle $ $(i,j=1,2,\cdots,N)$, then the eigenequation of the 2D extended SSH model is
	\begin{equation}
		\label{eq5}
		H^{\mathrm{2D}}\vert \psi^{\mathrm{2D}}\rangle=H^{\mathrm{2D}}\vert \psi_i \rangle \otimes \vert \psi_j \rangle =(E_i+E_j) \vert \psi_i \rangle \otimes \vert \psi_j \rangle.
	\end{equation}
	Because the wave functions $\vert\psi_i\rangle$ and $\vert\psi_j\rangle$ of the edge states for the 1D extended SSH chain in Fig.~\ref{Fig-1}A can localize at $\gamma$ sites per edge, the wave functions $\vert \psi^{\mathrm{2D}}\rangle$ of the corner states for the 2D extended SSH model can localize at $\gamma^2=2^2$ sites per corner, and four corners give 16 states. The corners states decay exponentially into the bulk. The further analysis about the topological properties of the 2D extended SSH model can be found in Supplementary Text \ref{z3}.
	
	Next, we construct a new class of higher-order topological insulators -- 2D local aperiodic systems to show the localized 0D topological states inherited from the 0D corner states, shown in Fig.~\ref{Fig-4}D based on the 1D local aperiodic chain in Fig.~\ref{Fig-1}B, whose parameters are in Fig.~\ref{Fig-1}C. Every row or column of the 2D aperiodic model is a 1D aperiodic chain, so the Hamiltonian of the 2D local aperiodic model is
	\begin{equation}
		\label{eq6}
		H^{\mathrm{2D}}_{\mathrm{eff}}= H_{\mathrm{eff}}\otimes I+I\otimes H_{\mathrm{eff}}.
	\end{equation}
	The energy spectrum of the 2D local aperiodic model is the same as that of the 2D extended SSH model, which is shown in Fig.~\ref{Fig-4}C, and the eigenmodes $(V^{\dagger}\vert \psi_i \rangle)\otimes (V^{\dagger}\vert \psi_j \rangle)$ of $H^{\mathrm{2D}}_{\mathrm{eff}}$ remain orthogonal to each other. The 2D aperiodic model in Fig.~\ref{Fig-4}D also has sixteen 0D states as shown in Fig.~\ref{Fig-4}E, whose topological property is inherited from the corner states of the 2D extended SSH model and determined by local topological marker $\tilde{\gamma}(l)$. But the localization sites are changed into sites $(m,n)\in\{1,8,13,16\}\times \{1,8,13,16\}$, which correspond to the localization sites (i.e., sites 1, 8, 13 and 16, see Fig.~\ref{Fig-1}E) of the 1D aperiodic chain. The $\mathbb{Z}$-classified higher-order topological insulators proposed in this work, i.e., the 2D extended SSH model and the transformed 2D aperiodic model, are intrinsically different from those built from sublattice multipole moment operators \cite{Z-the,Z-exp} for the following reasons. The former are protected by the winding number (the local topological marker) in each direction, and do not require noncommutative reflection symmetries, and have only positive couplings in one unit cell; however, the latter are protected by the multipole chiral numbers, and require noncommutative reflection symmetries, where both positive and negative couplings are needed to form a $\pi$ flux per plaquette in one unit cell.
	
	We design a 2D practical structure -- 2D acoustic coupled-resonators system shown in Fig.~\ref{Fig-4}F, to realize the model in Fig.~\ref{Fig-4}D. The structure in Fig.~\ref{Fig-4}F can also be constructed by the 1D  structure in Fig.~\ref{Fig-1}F. By calculating the eigenfrequencies and eigenmodes of the acoustic structure through COMSOL Multiphysics (which are  given in Supplementary Text \ref{z1}), we find that the sound intensities of the sixteen modes near ``zero energy" (in the frequency range of 2862--2876 Hz) show obvious amplitudes at sixteen resonators in Fig.~\ref{Fig-4}F, which is the typical field-distribution pattern of localized 0D topological states.
	
\begin{figure}
		\centering
		\includegraphics[width=0.8\linewidth]{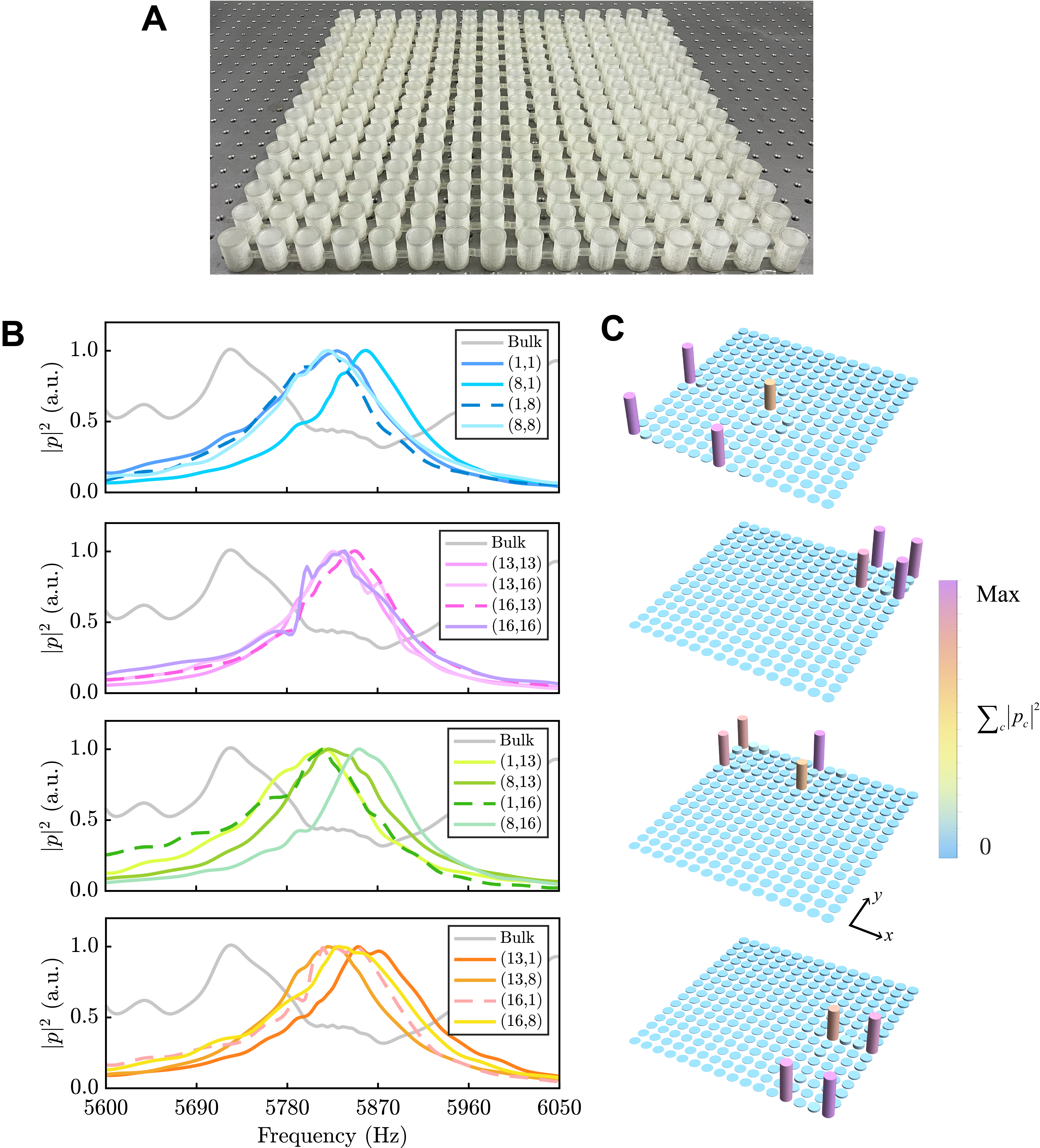}
		\caption{\label{Fig-5} \textbf{The fabricated 2D acoustic practical structure and the experimental results.} (\textbf{A}) The fabricated 2D acoustic coupled-resonators system. (\textbf{B}) The acoustic intensity response spectra of 0D topological states with excitations at sites $(m,n)\in\{1,8,13,16\}\times \{1,8,13,16\}$, where four sites are drawn in a group, and also nonlocalized bulk modes with excitation at site $(4,4)$ for comparison. (\textbf{C}) The sound field distributions of the sixteen localized 0D topological states with the same excitation sites as (B).}
	\end{figure}

	Finally, we fabricate the above 2D practical structure by three-dimensional stereolithography technique shown in Fig.~\ref{Fig-5}A (experimental details are given in Materials and Methods).
According to the dimensional analysis $\Pi_1=fL/c$ ($f$, $L$ and $c$ represent frequency, length and sound speed), we reduce the length to half of the original to save materials; then the corresponding frequency is expanded to twice the original one. First, a sound source is input from the side ports of the sites $(m,n)\in\{1,8,13,16\}\times\{1,8,13,16\}$ to excite the topological states, in the frequency range 5600--6050 Hz. Then, a microphone is used to measure the sound pressure at the upper ports of these sites, where the measurement sites are the same as the excitation sites as the usual practice~\cite{Z-exp,duotiao}. The response spectra of the sound intensity at these sixteen sites are shown in Fig.~\ref{Fig-5}B, where all response peaks are observed in the range 5815--5860 Hz (peak derivation of 100 Hz compared to the finite-element results 5728--5745 Hz given in Supplementary Text \ref{z2}, with an error about  $2\%$). Second, we measure the sound field distributions in the entire structure at these response peaks as shown in Fig.~\ref{Fig-5}C. The sixteen modes are localized at the sites $(m,n)\in\{1,8,13,16\}\times\{1,8,13,16\}$, showing typical characteristics of 0D topological states, which is consistent with the simulated results in Supplementary Text \ref{z2}. Although for the 2D local aperiodic model, the localized 0D topological states are embedded with the nonlocalized bulk modes, we can also successfully distinguish them. For comparison, we excite the bulk modes at site $(4,4)$ and also measure the sound pressure at site $(3,4)$. The bulk modes have no obvious peaks in the frequency range 5815--5860 Hz as the localized 0D topological states in Fig.~\ref{Fig-5}B. Thus, we demonstrate that the 2D higher-order local aperiodic model has sixteen localized 0D topological states.
	
	\subsection*{Discussion}
	Our work for the first time builds a correspondence between the nonlocal chiral-symmetric system and a nearest-neighbor-coupling chain with zero on-site potentials, thus providing a simple and effective way to demonstrate all the topological states of the nonlocal system experimentally. Based on the correspondence, we propose a class of local aperiodic chains that inherit the $\mathbb{Z}$-classified topological phases from the 1D nonlocal SSH chains, and realize the integer number of topological states in a 1D acoustic coupled-resonators system. Then, we construct a new class of higher-order topological insulators by the local aperiodic chains, which inherit the $\mathbb{Z}$-classified topological phases from the 2D nonlocal SSH model, and realize the 0D topological states in a 2D acoustic coupled-resonators system. These 0D topological states can be localized at arbitrary multiple positions in 1D and higher-order topological insulators, not only limited to the edges and corners. Although the topological insulators with winding number two in our work have four topological states in the 1D system and sixteen 0D topological states in the 2D system, the methods can be used to customize more 0D states with multi-position energy localization in multi-dimensional topological systems. Thus, the Householder transformation provides a powerful degree of freedom to reshape the wavefunction of topological states as required, so that the topologically protected bulk states can be more space-efficient \cite{Wang_2022,CABS-prl}. Such topological structures not only have applications to control sound emission \cite{emission,degenerate-disclination,duotiao}, but can also enhance sound intensity at multiple spacial positions so as to improve the sensitivity of sensor arrays in the fields of detection \cite{Mostafa1998} and imaging \cite{Epifanio_1999,computational-imaging}.
	The multiple 0D states at arbitrary positions enable tunable and programmable lasers and sasers \cite{ZHANG201976,zhang_low-threshold_2020,advs.201900401} by simply selecting the excitation sites, without altering the structure (Movies~S1--S3).
	
	\subsection*{Materials and Methods}
	\subsubsection*{Design of couplings in acoustic coupled-resonators system}\label{subsec3}
	
\begin{figure}
		\centering
		\includegraphics[width=0.9\linewidth]{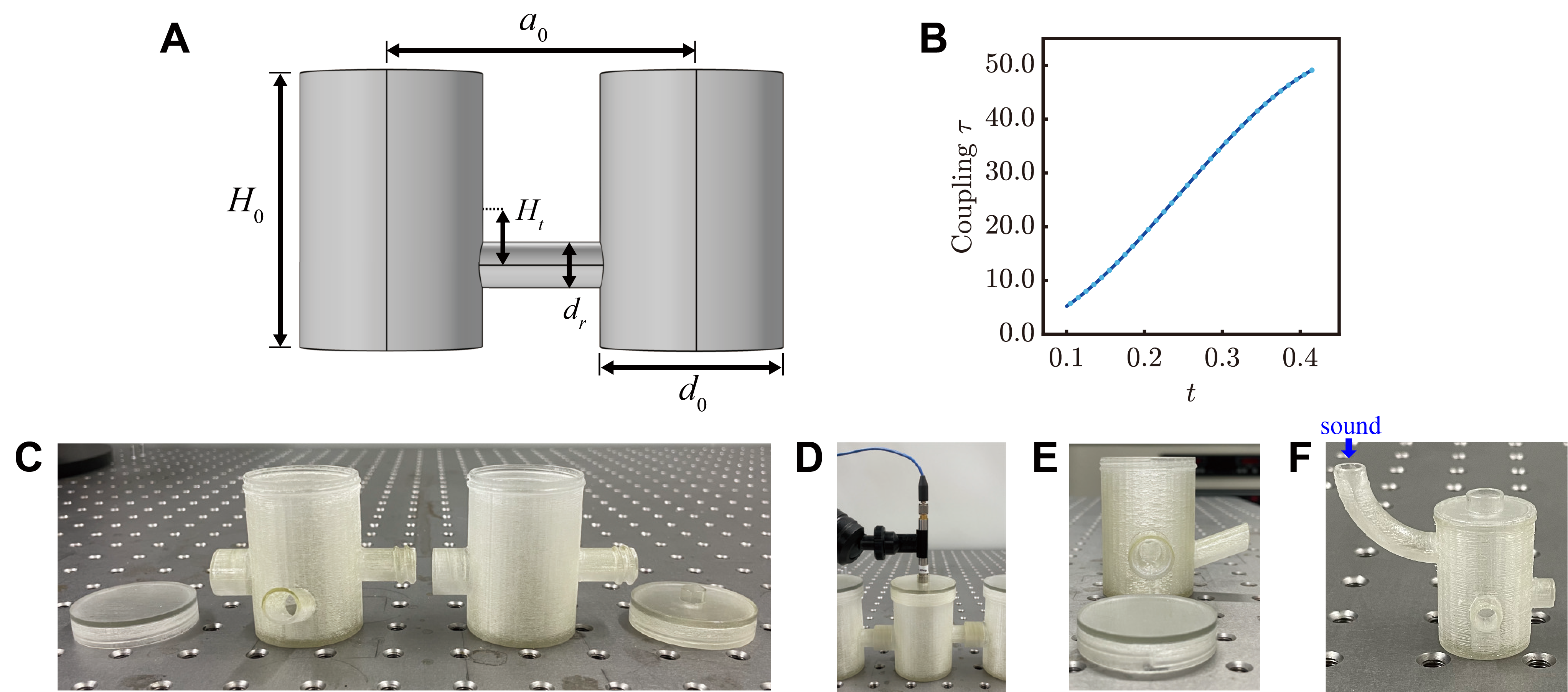}
		\caption{\label{Fig-7} \textbf{Construction of the coupling between two resonators and photographs of experimental parts.} (\textbf{A}) Two coupled resonators joined by a connecting tube. The height and diameter of a resonator are $H_0=60$ mm and $d_0=40$ mm, and the center-to-center distance of the two resonators is $a_0=66$ mm. The diameter of the connecting tube is $d_r=10$ mm, and the distance from the axis of a tube to the horizontal symmetry plane of resonators is $H_t$ ($H_t=t\cdot H_0$). (\textbf{B}) The relation between coupling $\tau$ and parameter $t$.  (\textbf{C}) 3D printed samples for assembly in 1D aperiodic structure. The main cavity of a part (i.e., a resonator with two attached connecting tubes) is on the middle right; and the main cavity of a part with a front opening port is on the middle left. The closed lid is on the leftmost side, and the lid with a upper port is on the rightmost side. (\textbf{D}) Schematic diagram of the resonator with a upper port for inserting microphone. (\textbf{E}) Side view of (C). (\textbf{F}) The designed curved tube on the side of the resonator for giving sound sources to excite 0D topological states in the 2D aperiodic structure.}
	\end{figure}

	We modulate the hopping amplitudes in the acoustic coupled-resonators system in Fig.~\ref{Fig-1}F by changing the vertical positions of the connecting tubes between every two coupled resonators. The height of each resonator is $H_0$ shown in Fig.~\ref{Fig-7}A, and $H_t$ represents the distance from the axis of a connecting tube to the horizontal symmetry plane of resonators.
	
	The relation between the hopping amplitude $\tau$ and distance $H_t$ can be determined as follows. For every two coupled resonators, the eigenequation is
	\begin{equation}
		\label{eqe:1}
		\begin{bmatrix} \omega_0&\tau(t)\\\tau(t)&\omega_0 \end{bmatrix}	\begin{bmatrix} \phi_1\\ \phi_2 \end{bmatrix}=\omega\begin{bmatrix} \phi_1\\ \phi_2 \end{bmatrix},
	\end{equation}
	where $\omega_0=v_c/(2H_0)$ is the eigenfrequency of each individual resonator, $v_c$ is the speed of sound; and $\tau(t)$ is the hopping amplitude, where $t$ is defined as $t\equiv H_t/H_0$. By solving Eq.~\ref{eqe:1}, the eigenfrequencies of the coupled resonators are $\omega_{\pm}=\omega_0\pm \tau(t)$. Then, the two coupled resonators are simulated by the finite-element software COMSOL Multiphysics to obtain the eigenfrequencies $\omega_{\pm}$, and the hopping amplitude is calculated by $\tau=(\omega_+-\omega_-)/2$. Thus, the relation between the hopping amplitude $\tau$ and $H_t~ ({\rm or}~t)$ is obtained as shown in Fig.~\ref{Fig-7}B.
	
	By using the above relation in Fig.~\ref{Fig-7}B, according to the coupling amplitudes for the mapped aperiodic chain shown in Fig.~\ref{Fig-1}C, the corresponding geometric distances $H_t$ for the acoustic coupled-resonators system are listed in Table~\ref{tab:table2}.
	\begin{table}
		\caption{\label{tab:table2} \textbf{Distances $H_t$ of the acoustic coupled-resonators system in Fig.~\ref{Fig-1}(\textbf{F}) from left to right, and the corresponding coupling amplitudes $\tau$ shown in Fig.~\ref{Fig-1}(\textbf{C}).}}
		\centering
		\begin{tabular}{ccccccccc}
			\hline
			$\tau$ & 16.971 & 42.954 & 26.244 & 43.334 & 25.108 & 42.362 & 12.982 & 6.776\\
			\hline
			$H_t~(\mathrm{mm})$ & 11.343 & 21.364 & 14.763 & 21.545 & 14.352 & 21.087 & 9.750 & 6.873\\
			\hline
			$\tau$ & 28.658 & 5.977 & 24.614 & 6.850 & 8.166 & 26.618 & 8.839 & {}\\
			\hline
			$H_t~(\mathrm{mm})$ & 15.641 & 6.438 & 14.174 & 6.913 & 7.585 & 14.898 & 7.913 & {}\\
			\hline
		\end{tabular}
	\end{table}
	
	\subsubsection*{Experimental specification}\label{subsec4}
	\paragraph*{1D practical structure.}
	We fabricated the 1D acoustic coupled-resonators system  consisting  of sixteen resonators joined by connecting tubes. Each part, i.e., one  resonator with two connecting tubes attached to it, was printed by a 3D printer (OBJET350 CONNEX3), using the high-stiffness material VeroClear. The thicknesses of the resonators and connecting tubes were 2~mm to mimic hard walls. Different parts were connected in sequence by interlocking screw threads.
	Two types of parts were used in the experiment: one had a closed upper lid, i.e. the left one in Fig.~\ref{Fig-7}C; the other one had a partly closed lid with a small opening of a radius 3.5 mm, i.e. the right one in Fig.~\ref{Fig-7}C. A 1/4 inch microphone was inserted into the opening of the lid, shown in Fig.~\ref{Fig-7}D, to measure the pressure field in the resonator. When a microphone was used to measure the pressure in a specific resonator, other resonators were all closed.
	
	In order to measure the acoustic intensity response spectra, a sound source was given on the leftmost side of the structure in Fig.~\ref{Fig-2}A; that is, the arbitrary waveform generator (Agilent 33522A) output an electrical signal, and then emitted an acoustic signal through the loudspeaker. We used the waveform generator to generate a single-frequency sine wave with a fixed amplitude, and the frequency was between 2790 Hz to 2980 Hz. Then, the microphone (PCB Piezotronics 378A14) was used to measure the sound pressure at sites 1, 8, 13 and 16 sequentially. Before the measurement, the microphone was calibrated by a sound level calibrator (Larson Davis CAL200) with an adapter (PCB Piezotronics ADP109). Finally, the microphone was connected to the signal conditioner (the Modal Shop 485B39) which was input into the computer to acquire the data by the SignalPad software.
	
	To measure the spatial sound field distribution, we gave in-phase and anti-phase excitation sound sources separately at sites 1 and 8, and sites 13 and 16. The excitation sources were input into the main cavities through the opening ports on their walls; Fig.~\ref{Fig-7}C shows the front view of such an opening port,  and Fig.~\ref{Fig-7}E shows the side view. The assembled overall systems are shown in Fig.~\ref{Fig-3}A and B. The sound sources were excited through loudspeakers. Two loudspeakers were separately connected to the two output channels of the waveform generator, and the two channels of the waveform generator were given signals with the same amplitude and frequency, but with the same phase or opposite phases with a difference 0 or $\pi$. Then, the absolute sound pressures at sixteen sites were measured with the microphone.
	
	\paragraph*{2D practical structure.}
	We fabricated the 2D acoustic coupled-resonators system consisting of $16\times 16$ resonators joined by connecting tubes. The entire system was 3D printed in four parts and finally spliced together, also in high-stiffness material VeroClear. To give sound sources at designated sites [sites $(m,n)\in\{1,8,13,16\}\times \{1,8,13,16\}$ and site $(4,4)$] in the 2D system, we designed an curved tube on the side of the resonator as shown in Fig.~\ref{Fig-7}F; when a certain site was selected for excitation, the resonator corresponding to this site in Fig.~\ref{Fig-5}A was replaced by the one with curved tube in Fig.~\ref{Fig-7}F, and the other resonators remained unchanged; then a loudspeaker was used at the upper end of the curved tube to give a sound source, and a support was also designed for the loudspeaker to keep it in a stable position. Each resonator was equipped with two types of lids: a closed lid and a partly closed lid with a upper port. When measuring the acoustic intensity response spectra, we used a loudspeaker to give the designated sites sound sources, and then placed a microphone at the excitation site [for excitations at sites $(m,n)\in\{1,8,13,16\}\times \{1,8,13,16\}$] or site $(3,4)$ [for excitation at site $(4,4)$] to measure sound pressure. Partly closed lids were selected for the sites for measurement, and closed lids for other sites. When measuring spatial sound field distribution, we placed a microphone at each resonator of the entire structure to measure the sound pressure.  The instrument models used in the 2D experiment were exactly the same as those in the 1D experiment.

	\nocite{unique,localizationdegree}
	%

	\section*{Acknowledgments}
	The authors would like to thank Professors Pu Chen, Shaoqiang Tang and Zifeng Yuan of Peking University for helpful discussions. We also thank Professor Faxin Li, and Dr Hao Qiu for help in experiments. 
	\paragraph{Funding:} This project was supported by the National Natural Science Foundation of China (Grants No.\ 11991033, No.\ 11890681, and No.\ 12232001).
	\paragraph{Author contributions:}
	Conceptualization: L.W., H.D., J.W.
	Methodology: Y.S., L.W., H.D., J.W.
	Investigation: Y.S., H.D., J.W.
	Supervision: L.W., J.W.
	Writing -- original draft: Y.S., L.W. 
	Writing -- review \& editing: Y.S., L.W., J.W.
	\paragraph{Competing Interests:} The authors declare no competing interests.
	\paragraph{Data and materials availability:} All data needed to evaluate the conclusions in the paper are present in the manuscript or the supplementary materials.

	\clearpage
	\section*{Supplementary Materials}
	\subsubsection*{This PDF file includes:}
	Supplementary Text\\
	Figs. S1 to S6\\
	Table S1\\
	Legend for movies S1 to S3\\
	References (\textit{18, 40, 42--45, 47, 50, 61, 62})
	\subsubsection*{Other Supplementary Materials for this manuscript includes the following:}
	Movies S1 to S3

	\newpage
	
	
	\renewcommand{\thefigure}{S\arabic{figure}}
	\renewcommand{\thetable}{S\arabic{table}}
	\renewcommand{\theequation}{S\arabic{equation}}
	\renewcommand{\thepage}{S\arabic{page}}
	\setcounter{figure}{0}
	\setcounter{table}{0}
	\setcounter{equation}{0}
	\setcounter{page}{1}
	

\begin{center}
	
	\section*{Supplementary Materials for\\[12pt]
		\Large \bf A new class of higher-order topological insulators that localize energy at arbitrary multiple sites}


	Yimeng Sun,$^{1}$ Linjuan Wang,$^{2\ast}$ Huiling Duan,$^{1}$ Jianxiang Wang$^{1\ast}$\\[10pt]
		\normalsize{$^{1}$State Key Laboratory for Turbulence and Complex Systems,}\\
		\normalsize{Department of Mechanics and Engineering Science,}\\
		\normalsize{College of Engineering, Peking University, Beijing 100871, China}\\
		\normalsize{$^{2}$School of Astronautics, Beihang University, Beijing 100191, China}\\[10pt]
		\normalsize{$^\ast$ Corresponding author. Email: wanglj@buaa.edu.cn; jxwang@pku.edu.cn}
		
	\end{center}
	
	\subsection*{This PDF file includes:}
	Supplementary Text\\
	Figs. S1 to S6\\
	Table S1\\
	Legends for movies S1 to S3\\
	References (\textit{18, 40, 42--45, 47, 50, 61, 62})
	\subsection*{Other Supplementary Materials for this manuscript includes the following:}
	Movies S1 to S3
	
	\clearpage
	
	\section*{Supplementary Text}
	\counterwithout{subsection}{section}
\subsection{Topological characterization}\label{a}
\begin{figure}[tbp]
	\centering
	\includegraphics[width=0.9\linewidth]{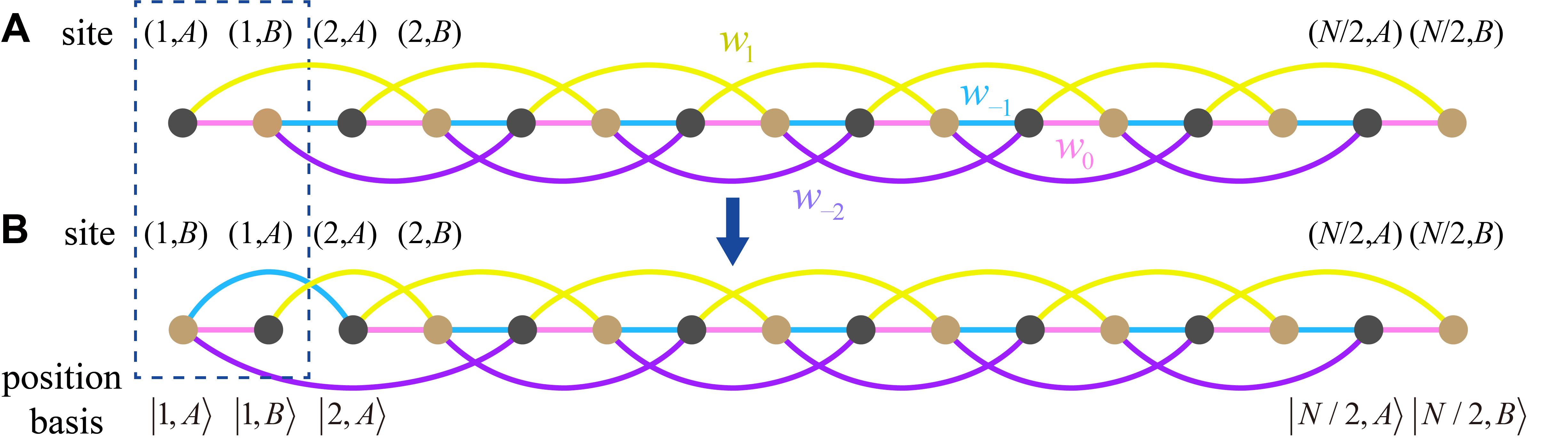}
	\caption{\label{Fig-sm1} \textbf{The anchor site for the improved Householder method.} (\textbf{A}) The original extended SSH chain. The default anchor site is the first site. The position basis is in the order of $\lvert 1,A\rangle, \lvert 1,B\rangle, \cdots, \lvert N/2,A\rangle, \lvert N/2,B\rangle$. The hoppings between sites are drawn as bonds (one color represents a hopping amplitude). (\textbf{B}) New chain constructed by ex\textbf{}changing the positions of sites $(1,A)$ and $(1,B)$ in the original chain, along with their respective dynamics (the hopping ``bonds''). This makes site $(1,B)$ -- now at the leftmost position $\lvert 1,A\rangle$ -- the new anchor site. Therefore, the dynamics of the second site in the original chain is equivalent to that of the first site in the new chain to be tridiagonalized by the Householder method.}
\end{figure}

For the bipartite extended SSH chain (in Fig.~\ref{Fig-1}, also in Fig.~\ref{Fig-sm1}A)  under the periodic boundary condition, the Hamiltonian is
\begin{equation}
\label{eqa:1}
{H} = \begin{pmatrix}
	0 & \sum\limits_{n=-n_\mathrm{left}}^{n_\mathrm{right}} w_n \mathrm{e}^{\mathrm{i}nk}
	\\\sum\limits_{n=-n_\mathrm{left}}^{n_\mathrm{right}} w_n \mathrm{e}^{-\mathrm{i}nk} & 0
\end{pmatrix} = d_x\cdot\sigma_x+d_y\cdot\sigma_y,
\end{equation}
where $\sigma_x$ and $\sigma_y$ are the Pauli matrices so that $d_x+\mathrm{i}d_y=\sum\limits_{n=-n_\mathrm{left}}^{n_\mathrm{right}} w_n \mathrm{e}^{-\mathrm{i}nk}$,
and $k$ is the Bloch wave number of the bipartite extended SSH chain. Then, we can show that the extended SSH chain has chiral symmetry (\textit{42}) by
\begin{equation}
\label{eqa:2}
{H}{\Gamma} =-{\Gamma}{H},
\end{equation}
where ${\Gamma}=\sigma_z$ is the chiral operator for a chain with two sites in a unit cell. Under the chiral symmetry, we should use the winding number to make a complete classification of the bandgap's topological property. The definition of the winding number (\textit{40, 42, 43}) is
\begin{equation}
\label{eqa:3}
\gamma=\frac{1}{2\pi \mathrm{i}}\int\nolimits_{-\pi}^{\pi} \mathrm{d}k \frac{\mathrm{d}}{\mathrm{d}k}\log{h(k)},
\end{equation}  where $ h(k)=\sum\limits_{n=-n_\mathrm{left}}^{n_\mathrm{right}} w_n \mathrm{e}^{-\mathrm{i}nk}=d_x(k)+\mathrm{i}d_y(k)$ and $\log{h(k)}=\log{\lvert h(k) \rvert }+\mathrm{i}\arg{h(k)}$, so
\begin{equation}
\label{eqa:4}
\gamma=\frac{1}{2\pi}\arg{h(k)} \rvert_{-\pi}^{\pi},
\end{equation}
where the branch cut of the logarithm is always shifted. The real part $d_x$ and the imaginary part $d_y$ of $h(k)$ can form a parametric curve $(d_x,d_y)$ about the wave number $k$. Eq.~\eqref{eqa:4} shows that the value of the winding number is determined by the number of times that the parametric curve turns anticlockwise around the origin as $k$ increases.
The winding number and the corresponding range of hopping amplitudes for the extended SSH chain in Fig.~\ref{Fig-sm1}A are given in Table~\ref{tab:table1}.
\begin{table}[tbp]
	\caption{\label{tab:table1} \textbf{The winding number of the bipartite extended SSH chain $(-2\leq n\leq 1)$ and the corresponding range of hopping amplitudes.} The quantities in the table are $x=w_{-1}/w_0$, $y=w_1/w_0$ and $z=w_{-2}/w_0$, where $C_1(x,y,z)=1+x+y+z$, $C_2(x,y,z)=1-x-y+z$, $C_3=z-z^2-xy+y^2$ and $C_4=\lvert y-x \rvert -2 \lvert z \rvert$. }
	\centering
	\begin{tabular}{cc}
		\hline
		$\lvert \gamma \rvert$ & Conditions\\
		\hline
		0 & $(C_3>0$ and $C_2>0)$ or $(C_3<0$, $C_2>0$ and $C_4>0)$\\
		1 & $C_2<0$ (i.e., $w_0+w_{-2}<w_{-1}+w_1$) \\
		2 & $C_3<0$ and $C_2>0$ and $C_4\leq 0$ \\
		\hline
	\end{tabular}
\end{table}

When the bipartite extended SSH chain is not boundless but finite under the open boundary condition, since the winding number is defined in the wave vector space rather than the real space, the topological property can be characterized by the local topological marker (LTM) (\textit{18}) in the real space:
\begin{equation}
\small
\label{eqa:5}
\gamma(l)=\frac{1}{2} \sum\limits_{a=A,B} \{(Q_{BA}[X,Q_{AB}])_{la,la} + (Q_{AB}[Q_{BA},X])_{la,la} \},
\end{equation}
where $Q_{AB}=\Gamma_{A} Q \Gamma_{B}$ and $Q_{BA}=\Gamma_{B} Q \Gamma_{A}$. $\Gamma_{A}$ ($\Gamma_{B}$) is defined as the projector onto the sublattice $A$ ($B$), and $Q$ is the ``flat-band Hamiltonian'' defined as $Q=P_+-P_-$, where $P_{+}=\psi_{+}\psi_{+}^{\dagger}$ and $P_{-}=\psi_{-}\psi_{-}^{\dagger}$ are the projectors of the energy spectrum above and below the bandgap, respectively. Here $\psi_{-}=[\psi_1,\psi_2,\cdots,\psi_{N/2}]$ and $\psi_{+}=[\psi_{N/2+1},\psi_{N/2+2},\cdots,\psi_{N}]$, where all eigenvectors are arranged so that the corresponding eigenvalues are in an ascending order; and the position operator is
\begin{equation}
\label{eqa:6}
\hat{X}=\sum\limits_{l=1}^{N/2}\sum\limits_{a=A,B} l \lvert l,a \rangle \langle l,a \rvert ,
\end{equation}
where $N$ is the total number of sites, and $\lvert l,a \rangle (a\in \{A,B\})$ represents the state of chain where the electron is at sublattice $a$ in the $l$th unit cell.

When unit cell $l$ is away from the edges of the finite chain, LTM $\gamma(l)$ is close to the value of the winding number, that is, when $l\rightarrow N/4+1$ and $N\rightarrow \infty$ (enough number of unit cells), then $\gamma(l)\rightarrow \gamma$.

\subsection{Improved Householder tridiagonalization and the anchor site}\label{b}
The Householder method can transform a general Hermitian (including real symmetric) matrix into a tridiagonal form, which can be further transformed by elementary row and column transformations so that the subdiagonal elements are positive, while the main-diagonal elements remain unchanged (the tridiagonalization begins from the first row and column of the symmetric matrix). Compared to the Lanczos tridiagonalization~(\textit{44, 45}), the Householder method has numerical stability when the dimension of the matrix increases. Here we use this method to map a topological system (Hamiltonian $H$) onto an aperiodic chain with only positive nearest-neighbor couplings (tridiagonal Hamiltonian $H_{\mathrm{eff}}$):
\begin{equation}
\label{eqb:1}
V^{\dagger}HV=H_{\mathrm{eff}}=\tilde{H},
\end{equation}
where $V$ is a unitary matrix which meets $VV^{\dagger}=I$, and the elements in matrix $H_{\mathrm{eff}}$ are all real [because the Householder method specifies that the subdiagonal elements are real numbers, and from Eq.~\eqref{eqb:1}, $H_{\mathrm{eff}}$ after transformation is also a Hermitian matrix, which ensures the diagonal elements are real numbers too]. According to
\begin{equation}
\label{eqb:2}
(V^{\dagger}HV)(V^{\dagger}\psi)=E(V^{\dagger}\psi),
\end{equation}
wave function $\psi$ will transform into $\tilde{\psi}=V^{\dagger}\psi$.

In the original Householder method (\textit{47}), $V_{11}=1$ $(V\neq V^{\dagger})$, then
\begin{equation}
\label{eqb:3}
\tilde{\psi}=V^{\dagger}\psi=\begin{bmatrix} 1&0\\0&A \end{bmatrix}^{\dagger}\begin{bmatrix} \phi_1\\ \phi_{\mathrm{else}} \end{bmatrix}=\begin{bmatrix} \phi_1\\ A\phi_{\mathrm{else}} \end{bmatrix},
\end{equation}
which means that the dynamics of the first site in the original system is faithfully mapped to that of the first site in the aperiodic chain ($\phi_m$ is the $m$th component of $\psi$). Here, we make such an improvement that \emph{the dynamics of any chosen site in the original system can also be faithfully mapped by changing the anchor site} as follows. When the first site and any other site (i.e., the $m$th site) in the original system, with their respective dynamical properties, exchange their positions (the process, with the extended SSH chain as the original system, is shown in Fig.~\ref{Fig-sm1} with $m=2$), it is equivalent to exchanging the first and $m$th rows, and the corresponding two columns of the original Hamiltonian $H$, using permutation matrix $\bar{V}=\bar{V}^{\dagger}$. Then we tridiagonalize the new Hamiltonian (i.e., $\bar{V}^{\dagger}H\bar{V}$) as
\begin{equation}
\label{eqb:4}
V^{*\dagger}(\bar{V}^{\dagger}H\bar{V})V^{*}=H^{*}_{\mathrm{eff}},
\end{equation}
where $V^{*}$ (remaining $V^{*}_{11}=1$) is different from $V$ in Eq.~\eqref{eqb:1} and $H^{*}_{\mathrm{eff}}$ is different from $H_{\mathrm{eff}}$, so that the dynamics of the $m$th site in the original system can be faithfully mapped. More specifically, Eq.~\eqref{eqb:4} is equivalent to
\begin{equation}
\label{eqb:5}
(\bar{V}V^{*})^{\dagger}H(\bar{V}V^{*})=H^{*}_{\mathrm{eff}},
\end{equation}
that is, our transformation enables $	(\bar{V}V^{*})_{m1}=1$. Then
\begin{equation}
\label{eqb:6}
\tilde{\psi}
=(\bar{V}V^*)^{\dagger}\psi=\begin{bmatrix} 0&B\\1&0\\0&C \end{bmatrix}^{\dagger}	\begin{bmatrix} \phi_{\mathrm{u}}\\\phi_m\\\phi_{\mathrm{d}} \end{bmatrix}
=\begin{bmatrix} 0&1&0\\B&0&C \end{bmatrix}\begin{bmatrix} \phi_{\mathrm{u}}\\\phi_m\\\phi_{\mathrm{d}} \end{bmatrix}=\begin{bmatrix} \phi_m\\B\phi_{\mathrm{u}}+C\phi_{\mathrm{d}} \end{bmatrix},
\end{equation}
which also means that the dynamics of the $m$th site in the original system is faithfully mapped to the first site in the aperiodic chain.

In the end of this section, we give the Matlab codes for constructing unitary matrix $V$ and tridiagonal matrix $H_{\mathrm{eff}}$ by the improved Householder method using Hermitian matrix $H$ as follows:
\begin{quote}
\begin{small}
	\begin{verbatim}
		function [V, Heff] = Householder(H)
		if ishermitian(H)
		if isreal(H)
		[V, Heff] = HouseholderReal(H);
		else
		[V, Heff] = HouseholderComplex(H);
		end
		else
		error('Invalid input')
		end
		end
		
		function [V, Heff] = HouseholderReal(H)
		n = length(H);
		V = eye(n);
		for k = 1:n-1
		x = H(k+1:n,k);
		if norm(x,1) < eps
		continue
		end
		s = norm(x);
		u = zeros([n 1]);
		u(k+1:n) = x;
		if x(1) >= 0
		u(k+1) = u(k+1) + s;
		else
		u(k+1) = u(k+1) - s;
		end
		h = norm(u)^2 / 2;
		P = eye(n) - u * u' / h;
		V = V * P;
		H = P * H * P;
		end
		Heff = diag(diag(H)) + diag(diag(H,-1),-1) ...
		+ diag(diag(H,-1),1);
		end	
		
		function [V, Heff] = HouseholderComplex(H)
		n = length(H);
		V = eye(n);
		for k = 1:n-1
		x = H(k+1:n,k);
		if norm(x,1) < eps
		continue
		end
		s = norm(x);
		u = zeros([n 1]);
		u(k+1:n) = x;
		if real(x(1)) >= 0
		u(k+1) = u(k+1) + s;
		else
		u(k+1) = u(k+1) - s;
		end
		h = norm(u)^2 / 2 * (1 - imag(u(k+1))/real(u(k+1)) * 1i);
		P = eye(n) - u * u' / h;
		V = V * P;
		H = P' * H * P;
		end
		Heff = diag(diag(H)) + diag(diag(H,-1),-1) ...
		+ diag(conj(diag(H,-1)),1);
		end
	\end{verbatim}
\end{small}
\end{quote}
The Householder transformation each time annihilates the nontridiagonal elements of its $k$th row and column, and repeats $n-1$ times to finally obtain a tridiagonal matrix. From the above codes, $P_k~(k=1,2,\cdots,n-1)$ is defined as the Householder reflector at the $k$th step which zeros the nontridiagonal elements of the $k$th row and column of matrix $H$, so the unitary matrix is $V=P_1P_2\cdots P_{n-1}$.

In addition, when the $m$th site in the original system is chosen as the anchor site, we also give the Matlab codes for constructing $\bar{V}V^{*}$ (i.e., the final $V$ in the code) and $H_{\mathrm{eff}}$ (also making all subdiagonal elements of $H_{\mathrm{eff}}$ positive real numbers) as follows:
\begin{quote}
\begin{small}
	\begin{verbatim}
		% Before used, Hamiltonian matrix H
		% and anchor site number m are needed
		t = length(H)/2;
		newAnchor = m;
		if newAnchor == 1
		[V, Heff] = Householder(H);
		elseif newAnchor <= 2*t
		[V, Heff] = Householder(H([newAnchor, 2:newAnchor-1, 1, ...
		newAnchor+1:end], [newAnchor, 2:newAnchor-1, 1, ...
		newAnchor+1:end]));
		V = V([newAnchor, 2:newAnchor-1, 1, newAnchor+1:end], :);
		else
		error('Invalid newAnchor number');
		end
		% The following making all subdiagonal elements
		% of Heff positive real numbers, but not
		% changing the signs of main-diagonal elements
		for ii = 1:2*t-1
		if Heff(ii,ii+1) < 0
		Heff(ii,ii+1) = -Heff(ii,ii+1);
		Heff(ii+1,ii) = -Heff(ii+1,ii);
		V(:,ii+1:end) = -V(:,ii+1:end);
		end
		end
	\end{verbatim}
\end{small}
\end{quote}

\subsection{Derivation of properties of the Householder method}\label{c}
In this section, we derive the following lemma:\begin{lemma}
\label{Theorem-1}
For a Hermitian matrix $H_{n\times n}$, the tridiagonalized matrix by the Householder method remains the same, when simultaneously exchanging any two rows and the corresponding two columns, except for the first row and first column, of $H$.	\end{lemma}The real symmetric matrix is used for derivation, although the same idea also applies to Hermitian matrices. We consider the nontrivial case that the tridiagonalized matrix has nonzero subdiagonal elements. For instance, a block matrix
\begin{equation}
\label{eqh:1}
H^{(0)}_{n\times n}=\begin{bmatrix} a_{11}&x^{(0)T}_{1\times (n-1)}\\x^{(0)}_{(n-1)\times 1}& B^{(0)}_{(n-1)\times (n-1)} \end{bmatrix}
\end{equation}
is used for tridiagonalization by the Householder method, and finally transformed as
\begin{equation}
\label{eqh:2}
H_{\mathrm{eff}}=\begin{bmatrix} d_1&e_1& & & & \\e_1&d_2&e_2& & &\\ &e_2&d_3& & &\\ & &\ddots&\ddots&\ddots&\\ & & & & d_{n-1}&e_{n-1}\\ & & & & e_{n-1}&d_n
\end{bmatrix}_{n\times n}.
\end{equation}
In Eq.~\eqref{eqh:2},  $d_1=a_{11}$ and $\lvert e_1 \rvert=\lVert x^{(0)} \rVert$. Then, we aim to seek a simple method to derive the expressions for $d_m~(m=1,2,\cdots,n)$ and $\lvert e_m \rvert~(m=1,2,\cdots,n-1)$. To this end, instead of considering the Householder tridiagonalization of  $H^{(0)}_{n\times n}$ in (\ref{eqh:1}),
we consider the Householder tridiagonalization of the following augmented matrix $\mathring{H}^{(0)}_{(n+1)\times (n+1)}$ where a zero row and a zero column are inserted into $H^{(0)}_{n\times n}$ as
\begin{equation}
\label{eqh:3}
\mathring{H}^{(0)}_{(n+1)\times (n+1)}=\left[
\begin{array}{c|c|c} a_{11}&0&x^{(0)T}_{1\times (n-1)}\\\hline0&0&0\\\hline x^{(0)}_{(n-1)\times 1}&0& B^{(0)}_{(n-1)\times (n-1)} \end{array}
\right],
\end{equation} and then prove its
$d_2$ and $\lvert e_2 \rvert$ are the same as those of the Householder tridiagonalization of the original $H^{(0)}_{n\times n}$.

Firstly, $\mathring{H}^{(0)}_{(n+1)\times (n+1)}$ is subjected to one single Householder transformation which annihilates the nontridiagonal elements of only its first row and column. Therefore, $X^{(0)}_{n\times1}\equiv \begin{bmatrix}
0\\x^{(0)}_{(n-1)\times 1}
\end{bmatrix}$ is transformed into $\begin{bmatrix}
\lVert x^{(0)} \rVert\\0_{(n-1)\times 1}
\end{bmatrix}$ (let the sign in front of $\lVert x^{(0)} \rVert$ positive, which does not affect the results of $d_2$ and $\lvert e_2\rvert$); at the same time,
$A^{(0)}_{n\times n}\equiv \begin{bmatrix}
0&0_{1\times (n-1)}\\0_{(n-1)\times1}&B^{(0)}_{(n-1)\times (n-1)}
\end{bmatrix}$is transformed into
\begin{equation}
\label{eqh:32}
H^{(1)}_{n\times n}\equiv \begin{bmatrix}
	d_2&x^{(1)T}_{1\times (n-1)}\\
	x^{(1)}_{(n-1)\times 1}&B^{(1)}_{(n-1)\times (n-1)}
\end{bmatrix},
\end{equation}
so that the expressions of $d_2$ and $\lvert e_2 \rvert=\lVert x^{(1)}\rVert$ can be given by $x^{(0)}$ and $B^{(0)}$. To this end, we construct  $u^{(0)}_{n\times1}\equiv\begin{bmatrix}
\lVert x^{(0)} \rVert\\
x^{(0)}_{(n-1)\times 1}
\end{bmatrix}$  by $X^{(0)}_{n\times1}$. Then, we introduce
\begin{equation}
\label{eqh:4}
P^{(0)}_{n\times n}\equiv I-2\frac{u^{(0)}u^{(0)T}}{\lVert u^{(0)}\rVert^2}\\
=\begin{bmatrix}
	1&0\\0&I
\end{bmatrix}-\frac{2}{\lVert u^{(0)}\rVert^2}
\begin{bmatrix}
	\lVert x^{(0)} \rVert^2&\lVert x^{(0)} \rVert x^{(0)T}\\ \lVert x^{(0)} \rVert x^{(0)}&x^{(0)}x^{(0)T}
\end{bmatrix}.
\end{equation}
With the help of $P^{(0)}$ (got from $X^{(0)}$), $A^{(0)}$ is transformed into $H^{(1)}=P^{(0)T}A^{(0)}P^{(0)}$ (the blocks of $H^{(1)}$, $P^{(0)}$ and $A^{(0)}$ are partitioned into subblocks in the same forms), so
\begin{equation}
\label{eqh:5}
d_2=(P^{(0)T}A^{(0)}P^{(0)})_{11}
=P^{(0)T}_{12}A^{(0)}_{22}P^{(0)}_{21}
=\frac{x^{(0)T}B^{(0)}x^{(0)}}{\lVert x^{(0)}\rVert^2},
\end{equation}
\begin{equation}
\label{eqh:6}
x^{(1)}=(P^{(0)T}A^{(0)}P^{(0)})_{21}
=-\frac{B^{(0)}x^{(0)}}{\lVert x^{(0)}\rVert}+\frac{x^{(0)}x^{(0)T}B^{(0)}x^{(0)}}{\lVert x^{(0)}\rVert^3},
\end{equation}
\begin{equation}
\label{eqh:7}
e_2=\pm\lVert x^{(1)}\rVert,
\end{equation}
and
\begin{equation}
\label{eqh:8}
B^{(1)}=(P^{(0)T}A^{(0)}P^{(0)})_{22}
=\left(I-\frac{x^{(0)}x^{(0)T}}{\lVert x^{(0)}\rVert^2}\right)B^{(0)}\left(I-\frac{x^{(0)}x^{(0)T}}{\lVert x^{(0)}\rVert^2}\right),
\end{equation}
where the indices in Eqs.~\eqref{eqh:5} - \eqref{eqh:8} correspond to the subblock matrices, that is, $A^{(0)}_{22}=B^{(0)}$, $P^{(0)}_{11}=1-\frac{2}{\lVert u^{(0)}\rVert^2}\lVert x^{(0)} \rVert^2$, $P^{(0)}_{12}=P^{(0)T}_{21}=-\frac{2}{\lVert u^{(0)}\rVert^2}\lVert x^{(0)} \rVert x^{(0)T}$ and $P^{(0)}_{22}=I-\frac{2}{\lVert u^{(0)}\rVert^2}x^{(0)}x^{(0)T}$.

Secondly, we prove that $H^{(0)}$ [in Eq.~\eqref{eqh:1} without inserting zeros] and $\mathring{H}^{(0)}$ [in Eq.~\eqref{eqh:3} inserting a zero row and a zero column] have the same  $d_2$ and $\lvert e_2\rvert$ after one single Householder transformation. At this time, $H^{(0)}$ is subjected to one single Householder transformation, and $B^{(0)}_{(n-1)\times(n-1)}$ is directly transformed by $x^{(0)}_{(n-1)\times 1}$; then we compare the new expressions of $d_2$ and $\lvert e_2\rvert$ with Eqs.~\eqref{eqh:5} and \eqref{eqh:7}. We define
\begin{equation}
\label{eqh:12}
u\equiv x^{(0)}+\lVert x^{(0)} \rVert {\mathrm {e}}_1
\end{equation}
by $x^{(0)}$ [where ${\mathrm {e}}_1=(1,0,\cdots,0)^T$], and $\lVert u\rVert^2=2(x^{(0)}_1+\lVert x^{(0)} \rVert) \lVert x^{(0)}  \rVert$. Then,
we introduce	\begin{equation}
\label{eqh:13}
P_{(n-1)\times (n-1)}\equiv I-2\frac{uu^T}{\lVert u\rVert^2},
\end{equation}
and then, $d_{2}$ can be expressed as
\begin{equation}
\label{eqh:14}
d_2=(PB^{(0)}P)_{11}=\sum_{i=1}^{n-1}\sum_{j=1}^{n-1}P_{1j}B^{(0)}_{ji}P_{i1}=\frac{x^{(0)T}B^{(0)}x^{(0)}}{\lVert x^{(0)}\rVert^2},
\end{equation}
where
\begin{equation}
\label{eqh:15}
\begin{split}
	P_{1j}&=\delta_{1j}-\frac{2(x_1^{(0)}+\lVert x^{(0)}\rVert)(x_j^{(0)}+\lVert x^{(0)}\rVert\delta_{j1})}{2(x_1^{(0)}+\lVert x^{(0)}\rVert)\lVert x^{(0)}\rVert}\\
	&=\begin{cases}
		1-\frac{\lVert x^{(0)}\rVert+x_1^{(0)}}{\lVert x^{(0)}\rVert}=-\frac{x_1^{(0)}}{\lVert x^{(0)}\rVert}\quad (j=1) \\
		-\frac{x_j^{(0)}}{\lVert x^{(0)}\rVert}\quad (j\neq 1)
	\end{cases}
	=-\frac{x_j^{(0)}}{\lVert x^{(0)}\rVert}.
\end{split}
\end{equation}
$d_2$ given in Eq.~\eqref{eqh:14} is the same as Eq.~\eqref{eqh:5}.

Then, we derive $\lvert e_2\rvert$ by $H^{(0)}$ without inserting zeros. Let $y_m\equiv (PB^{(0)}P)_{m1}=P_{mi}B_{ij}^{(0)}P_{j1}  (m\geq 2)$, so $\lvert e_2\rvert=\lVert y \rVert=\sqrt{y^Ty}$. Considering Eq.~\eqref{eqh:15}, then
\begin{equation}
\small
\label{eqh:16}
\begin{split}
	y^Ty&=\sum_{m=2}^{n-1}\sum_{i=1}^{n-1}\sum_{i'=1}^{n-1}\frac{1}{\lVert x^{(0)}\rVert^2}(B^{(0)}x^{(0)})^T_{1i}P_{im}P_{mi'}(B^{(0)}x^{(0)})_{i'1}\\
	&=\frac{1}{\lVert x^{(0)}\rVert^2}(B^{(0)}x^{(0)})^T\left(I-\frac{x^{(0)}x^{(0)T}}{\lVert x^{(0)}\rVert^2}\right)(B^{(0)}x^{(0)}),
\end{split}
\end{equation}
where
\begin{equation}
\label{eqh:17}
\begin{split}
	P_{mi}(m\geq2)&=\delta_{mi}-\frac{x^{(0)}_m(x^{(0)}_i+\lVert x^{(0)}\rVert\delta_{1i})}{\lVert x^{(0)}\rVert(x^{(0)}_1+\lVert x^{(0)}\rVert)}
	=\begin{cases}
		-\frac{x^{(0)}_m}{\lVert x^{(0)}\rVert}\quad (i=1) \\
		\delta_{mi}-\frac{x^{(0)}_mx^{(0)}_i}{\lVert x^{(0)}\rVert(x^{(0)}_1+\lVert x^{(0)}\rVert)}\quad (i\neq 1)
	\end{cases},
\end{split}
\end{equation}
and
\begin{equation}
\label{eqh:18}
\begin{split}
	\sum\limits_{m=2}^{n-1}P_{im}P_{mi'}&=
	\begin{cases}
		\frac{\lVert x^{(0)}\rVert^2-(x^{(0)}_1)^2}{\lVert x^{(0)}\rVert^2}\quad (i=1,i'=1) \\
		-\frac{x^{(0)}_1x^{(0)}_{i'}}{\lVert x^{(0)}\rVert^2} \quad (i=1,i'\neq1) \\
		-\frac{x^{(0)}_1x^{(0)}_{i}}{\lVert x^{(0)}\rVert^2} \quad (i\neq1,i'=1)\\
		\delta_{ii'}-\frac{x^{(0)}_{i'}x^{(0)}_{i}}{\lVert x^{(0)}\rVert^2} \quad (i\neq1,i'\neq1)
	\end{cases}
	=\left(I-\frac{x^{(0)}x^{(0)T}}{\lVert x^{(0)}\rVert^2}\right)_{ii'}.
\end{split}
\end{equation}

From Eqs.~\eqref{eqh:6} and \eqref{eqh:7}, $x^{(1)T}x^{(1)}$ can be expressed as
\begin{equation}
\label{eqh:20}
\begin{split}
	x^{(1)T}x^{(1)}&=(B^{(0)}x^{(0)})^T\cdot\left(\frac{-\lVert x^{(0)}\rVert^2I+x^{(0)}x^{(0)T}}{\lVert x^{(0)}\rVert^3}\right)^T \cdot\left(\frac{-\lVert x^{(0)}\rVert^2I+x^{(0)}x^{(0)T}}{\lVert x^{(0)}\rVert^3}\right)\cdot(B^{(0)}x^{(0)})\\
	&=(B^{(0)}x^{(0)})^T\frac{1}{\lVert x^{(0)}\rVert^2}\left(\frac{\lVert x^{(0)}\rVert^2I-x^{(0)}x^{(0)T}}{\lVert x^{(0)}\rVert^2}\right)(B^{(0)}x^{(0)}),
\end{split}
\end{equation}
Eq.~\eqref{eqh:20} and Eq.~\eqref{eqh:16} show that the expressions of $\lvert e_2\rvert$ obtained by the two methods are the same.

Thirdly, we use the simple zero-insertion method to get the expressions of $d_m$ and $\lvert e_m \rvert$ by $x^{(0)}$ and $B^{(0)}$. According to the above derivation, $d_2$ and $\lvert e_2 \rvert$ are given by one single Householder transformation of $\mathring{H}^{(0)}$. Then, a zero row and a zero column are inserted into $H^{(1)}_{n\times n}$ [equation \eqref{eqh:32}] as
\begin{equation}
\label{eqh:31}
\mathring{H}^{(1)}_{(n+1)\times (n+1)}=\left[
\begin{array}{c|c|c} d_{2}&0&x^{(1)T}_{1\times (n-1)}\\\hline0&0&0\\\hline x^{(1)}_{(n-1)\times 1}&0& B^{(1)}_{(n-1)\times (n-1)} \end{array}
\right],
\end{equation}
and then, $d_3$ and $\lvert e_3 \rvert$ are given by one single Householder transformation of $\mathring{H}^{(1)}$ (at this time, $X^{(1)}_{n\times1}\equiv \begin{bmatrix}
0\\x^{(1)}_{(n-1)\times 1}
\end{bmatrix}$ is used to transform
$A^{(1)}_{n\times n}\equiv \begin{bmatrix}
0&0\\0&B^{(1)}_{(n-1)\times (n-1)}
\end{bmatrix}$). According to this recurrence relation [the recurrence formulas are given by changing $B^{(0)}$ and $x^{(0)}$ in Eqs.~\eqref{eqh:5} - \eqref{eqh:8} into $B^{(m)}$ and $x^{(m)}$, $B^{(1)}$ and $x^{(1)}$ into $B^{(m+1)}$ and $x^{(m+1)}$, $d_2$ and $e_2$ into $d_{m+2}$ and $e_{m+2}$], the recursion continues until $d_n$ and $\lvert e_n \rvert$ are obtained by one single Householder transformation of $\mathring{H}^{(n-2)}$ (at this time, $X^{(n-2)}_{n\times1}\equiv \begin{bmatrix}
0\\x^{(n-2)}_{(n-1)\times 1}
\end{bmatrix}$ is used to transform
$A^{(n-2)}_{n\times n}\equiv \begin{bmatrix}
0&0\\0&B^{(n-2)}_{(n-1)\times (n-1)}
\end{bmatrix}$),
where $e_n$ is equal to 0, and
\begin{equation}
\label{eqh:9}
H^{(n-1)}_{n\times n}=\begin{bmatrix}
	d_n&0_{1\times(n-1)}\\0_{(n-1)\times1}&0_{(n-1)\times(n-1)}
\end{bmatrix}.
\end{equation}
In the process of recursion, we apply the action of simultaneously inserting a zero row and a zero column for $n-1$ times in total. The process of recursion is summarized as
\[ \small \begin{split}
d_1, {x}^{(0)}(\lvert e_1 \rvert), B^{(0)} &\Longrightarrow d_2, x^{(1)}(\lvert e_2 \rvert), B^{(1)} \\
&\Longrightarrow d_3, x^{(2)}(\lvert e_3 \rvert), B^{(2)} \\
&\Longrightarrow \cdots \\
&\Longrightarrow d_{n-1}, x^{(n-2)}(\lvert e_{n-1} \rvert), B^{(n-2)} \\
&\Longrightarrow d_{n}, x^{(n-1)}=0~(\lvert e_{n} \rvert=0), B^{(n-1)}=0.
\end{split} \]
Through this recursion, all the elements $d_m~(m=2,\cdots,n)$ and $\lvert e_m\rvert~(m=1,\cdots,n-1)$ in Eq.~\eqref{eqh:2} can be expressed ultimately by $x^{(0)}$ and $B^{(0)}$. The advantage of the zero-insertion method is that the dimensions of the transformed vectors $x^{(m)}$ and matrices $B^{(m)}$ remain the same in the recursion at each time $(m=0,\cdots,n-1)$.

Finally, we use the simple recursion to prove Lemma~\ref{Theorem-1}. For the matrix $H$ to be tridiagonalized by the Householder method, when any two rows and the corresponding columns are exchanged (excluding the first row and column), $x^{(0)}$ and $B^{(0)}$ in Eq.~\eqref{eqh:1} become
\begin{equation}
\label{eqh:10}
x^{(0)}_{\mathrm{new}}=Q^Tx^{(0)},
\end{equation}
\begin{equation}
\label{eqh:11}
B^{(0)}_{\mathrm{new}}=Q^T B^{(0)}Q,
\end{equation}
where $Q$ is the corresponding permutation matrix. Substituting Eqs.~\eqref{eqh:10} and \eqref{eqh:11} into Eqs.~\eqref{eqh:5} - \eqref{eqh:8}, we find that $d_2$ and $\lvert e_2\rvert$ remain unchanged, but $x^{(1)}_{\mathrm{new}}=Q^T x^{(1)}$ and $B^{(1)}_{\mathrm{new}}=Q^T B^{(1)}Q$, which have the same forms as Eqs.~\eqref{eqh:10} and \eqref{eqh:11}. Similarly, according to the recurrence relation, we find that $d_m~(m=1,\cdots,n)$ and $\lvert e_m\rvert~(m=1,\cdots,n-1)$ remain unchanged as well.
Thus, Lemma~\ref{Theorem-1} is proved.

In virtue of Lemma~\ref{Theorem-1}, with the improved Householder method, the choice of the anchor site is divided into two cases to prove Theorem~\ref{Theorem-2}. For a chiral-symmetric system, we find a suitable position basis with two sublattices of type $A$ and type $B$, so that the Hamiltonian matrix satisfies
\begin{equation}
\label{eqh:21}
\Gamma^{\dagger}H\Gamma=-H,
\end{equation}
where $\Gamma=I_{(n/2)\times(n/2)}\otimes \sigma_z$ is the chiral operator.

First, when the anchor site is chosen as site $(l,A)$ [where $(l,A)$ represents the $l$th site at sublattice $A$], matrix $\bar{V}^{\dagger}H\bar{V}$ in Eq.~\eqref{eqb:4} is also chiral-symmetric:
\begin{equation}
\label{eqh:23}
\Gamma^{\dagger}(\bar{V}^{\dagger}H\bar{V})\Gamma=-(\bar{V}^{\dagger}H\bar{V}).
\end{equation}
Then, we get $\Gamma^{\dagger}P_k\Gamma=P_k~(k=1,2,\cdots,n-1)$ according to the construction of $P_k$ [where $P_k~(k=1,2,\cdots,n-1)$ is the Householder reflector at the $k$th single Householder transformation defined in Section~\ref{b}, which annihilates the nontridiagonal elements of the $k$th row and column of matrix $\bar{V}^{\dagger}H\bar{V}$]. Because $V^{*}$ in Eq.~\eqref{eqb:4} satisfies $V^{*}=P_1P_2\cdots P_{n-1}$, we obtain
\begin{equation}
\label{eqh:24}
\Gamma^{\dagger}V^*\Gamma=V^*.
\end{equation}
Substituting Eqs.~\eqref{eqh:23} and \eqref{eqh:24} into Eq.~\eqref{eqb:4}, we get $\Gamma^{\dagger}H_{\mathrm{eff}}^*\Gamma=-H_{\mathrm{eff}}^*$, so the on-site potentials of the mapped aperiodic chain with the anchor site at sublattice $A$ are all zero.

Second, when the anchor site is chosen as site $(l,B)$ [where $l$ represents the $(l,B)$th site at sublattice $B$], matrix $(\bar{V}^{\dagger}H\bar{V})_{n\times n}$ in Eq.~\eqref{eqb:4} is no longer chiral-symmetric, but we can selectively insert several zero rows and corresponding zero columns by introducing a matrix $C_{n\times n'}$ to make $[C^{\dagger}(\bar{V}^{\dagger}H\bar{V})C]_{n'\times n'}$ chiral-symmetric ($n'$ is also an even number).
It can be proved that the following matrices:
\begin{equation}
\label{eqh:26}
C_{n\times(n+2)}^{l=1}=\left[
\begin{array}{c|c|c|c}
	\boldsymbol{I}_{2\times2}&0_{2\times1}&0&0\\ \hline
	0&0&\boldsymbol{I}_{(n-2)\times(n-2)}&0_{(n-2)\times1}
\end{array}
\right],
\end{equation}
\begin{equation}
\label{eqh:27}
C_{n\times(n+2)}^{l=n/2}=\left[
\begin{array}{c|c|c|c|c}
	\boldsymbol{I}_{1\times1}&0_{1\times1}&0&0&0\\\hline
	0&0&\boldsymbol{I}_{(n-2)\times(n-2)}&0_{(n-2)\times1}&0\\\hline
	0&0&0&0&\boldsymbol{I}_{1\times1}
\end{array}
\right],
\end{equation}
\begin{equation}
\label{eqh:28}
\small	C_{n\times(n+4)}^{1<l<n/2}=\left[
\begin{array}{c|c|c|c|c|c|c|c}
	\boldsymbol{I}_{1\times1}&0_{1\times1}&0&0&0&0&0&0\\\hline 0&0 &\boldsymbol{I}_{(2l-2)\times(2l-2)}&0_{(2l-2)\times1}& 0&0 &0 &0 \\\hline0 & 0&0&0 &\boldsymbol{I}_{1\times1}&0_{1\times1}& 0& 0\\\hline0 &0 & 0&0 &0 &0 &\boldsymbol{I}_{(n-2l)\times(n-2l)}&0_{(n-2l)\times1}
\end{array}
\right]
\end{equation}
can make $[C^{\dagger}(\bar{V}^{\dagger}H\bar{V})C]_{n'\times n'}$ chiral-symmetric.
After introducing matrix $C$, by the Householder transformation:
\begin{equation}
\label{eqh:25}
V'^{\dagger}[C^{\dagger}(\bar{V}^{\dagger}H\bar{V})C]V'=H'_{\mathrm{eff}},
\end{equation}
where
\begin{equation}
\label{eqh:29}
\Gamma'^{\dagger}V'\Gamma'=V'\quad(\Gamma'=I_{(n'/2)\times(n'/2)}\otimes \sigma_z),
\end{equation}
we get $\Gamma'^{\dagger}H'_{\mathrm{eff}}\Gamma'=-H'_{\mathrm{eff}}$ (the main-diagonal elements of $H'_{\mathrm{eff}}$ are all zero). According to Lemma~\ref{Theorem-1}, $d_m~(m=1,\cdots,n)$ and $\lvert e_m\rvert~(m=1,\cdots,n-1)$ in $H'_{\mathrm{eff}}$ are the same as those in $H^{*}_{\mathrm{eff}}$ obtained by Eq.~\eqref{eqb:4}, and the last $(n'-n)$ rows and columns of $H'_{\mathrm{eff}}$ are all zero. Therefore, $H^{*}_{\mathrm{eff}}$ also satisfies $\Gamma^{\dagger}H^{*}_{\mathrm{eff}}\Gamma=-H^{*}_{\mathrm{eff}}$, the on-site potentials of the mapped aperiodic chain with the anchor site at sublattice $B$ are also all zero.

We have proved that all on-site potentials of the Householder-tridiagonalized chain mapped from any chiral-symmetric system are zero no matter which anchor site is chosen. Then, according to the reduction uniqueness theorem (\textit{61}), as long as the anchor site is fixed, the tridiagonalized matrix is unique no matter what tridiagonalization method is used in the proof. Thus, Theorem~\ref{Theorem-2} is proved.

Since $H_{\mathrm{eff}}$ in Eq.~\eqref{eqb:1} and $H_{\mathrm{eff}}^*$ in Eq.~\eqref{eqb:4} are all chiral-symmetric matrices, \emph{no matter which anchor site is chosen for the bipartite extended SSH chain, each pair of topological states for the Householder-tridiagonalized chain satisfy $\tilde{\psi}_2=\Gamma\tilde{\psi}_1$}, that is, the amplitude distributions are the same (i.e., $\lvert \langle m \vert \tilde{\psi_1}\rangle\rvert=\lvert \langle m \vert \tilde{\psi_2}\rangle\rvert$) but the phase distributions are different.

\subsection{Localization degree of topological state}\label{d}
For the original extended SSH chain with bulk winding number two under the open boundary conditions, if the topological state is localized at both the $p$th and $q$th sites near the edges, then after tridiagonalization, $\tilde{\psi}$ satisfies
\begin{equation}
\label{eqc:1}
\tilde{\psi}=V^{\dagger}\psi
=\begin{bmatrix} b_1&b_2&\cdots&b_N \end{bmatrix} \begin{bmatrix} \phi_1\\ \phi_2 \\ \vdots \\ \phi_N \end{bmatrix}=\phi_1[b_1]+\phi_2[b_2]+\cdots+\phi_N[b_N],
\end{equation}
where $N$ represents the total number of sites, and $[b_m]$ represents the $m$th column of matrix $V^{\dagger}$. As $\phi_p$ and $\phi_q$ are dominant components, $\tilde{\psi}\approx \phi_p[b_p]+\phi_q[b_q]$. Therefore, the spatial localization of topological states in the mapped aperiodic chain is not the same as that in the original chain, but depends on vectors $[b_p]$ and $[b_q]$. The inverse participation ratio (IPR) (\textit{45, 62}) of the eigenstate is introduced to quantify the degree of localization:
\begin{equation}
\label{eqc:2}
\textrm{IPR}(\lvert \psi \rangle)=\sum\limits_{m=1}^{N}\lvert \langle m\vert \psi \rangle \rvert^4,
\end{equation}
so for the state after tridiagonalization,
\begin{equation}
\label{eqc:3}
\textrm{IPR}(\lvert \tilde{\psi} \rangle)=\sum\limits_{m=1}^{N}\lvert \langle m\vert \tilde{\psi} \rangle \rvert^4=\sum\limits_{m=1}^{N}\lvert \langle m\vert V^{\dagger}\psi \rangle \rvert^4.
\end{equation}
More specifically, for the original extended SSH chain in Fig.~\ref{Fig-1}A (hopping amplitudes are $w_{-2}=33, w_{-1}=w_0=w_1=12$), a pair of topological states is localized at the first and sixteenth sites ($\textrm{IPR}=0.3271$), and the other pair of topological states is localized at the third and fourteenth sites ($\textrm{IPR}=0.3289$). Next, we analyze the degree of localization for the mapped aperiodic chain: the IPRs calculated by Eq.~\eqref{eqc:3} are 0.3268 and 0.3707 (close to the results of the original chain). Then, we trace the localization site of the aperiodic chain: corresponding to $\phi_1$, $\phi_3$, $\phi_{14}$ and $\phi_{16}$ of the original chain in Eq.~\eqref{eqc:1}, we get
\begin{equation}
\label{eqc:4}
\begin{split}
	[b_1]=[&\boldsymbol{\mathit{1}},0,0,0,0,0,0,0,0,0,0,0,0,0,0,0]^T,\\
	[b_3]=[&0,0,0.3951,0,-0.0267,0,-0.0520,0,0.1571,0,-0.1913,0,\boldsymbol{\mathit{0.7737}},0,-0.4250,0]^T,\\
	[b_{14}]=[&0,0,0,0,0,0.3370,0,0.3068,0,-0.0269,0,-0.0074,0,0.4063,0,\boldsymbol{\mathit{-0.7915}}]^T,\\
	[b_{16}]=[&0,0,0,0,0,0.0904,0,\boldsymbol{\mathit{0.9192}},0,-0.1042,0,-0.0039,0,-0.2529,0,0.2685]^T.
\end{split}
\end{equation}
The row number of the largest element in each of the column vectors in Eq.~\eqref{eqc:4} corresponds to the new localization site in the mapped aperiodic chain; the sites are 1, 8, 13 and 16.

Through the above analysis, the selection of an appropriate $V$ (or $\bar{V}V^*$, depending on anchor site) can make the topological states exhibit a high localization degree after the Householder transformation. 

Then, we attempt to give a more formal criterion regarding the inverse participation ratio (IPR) to define whether a particular state of local aperiodic chains is regarded as ``localized'' or not. For a $\mathbb{Z}$-classified SSH chain with long-range hoppings, according to the decay coefficient $\xi$ of the localized topological state and the number of localization sites, $n_\mathrm{l}$, we can define a standard reference IPR value for an exponentially decaying localized mode: $\mathrm{IPR}_\mathrm{st} = \frac{1 - \lvert \xi \rvert^2}{n_\mathrm{l} (1 + \lvert \xi \rvert^2)}$, as calculated below. When the IPR of a topological state in the transformed aperiodic system is greater than this standard $\mathrm{IPR}_\mathrm{st}$, we deem the topological state to be localized; otherwise not. The following are organized into two steps.

First, we consider the IPRs of exponentially localized states. For such a localized state which is localized at the first site, the wavefunction is expressed as $\psi = \sqrt{c}(1, \xi, \xi^2, \dots, \xi^{N-1})$, thus $\lvert\psi\rvert^2 = c(1, \xi^2, \xi^4, \dots, \xi^{2N-2})$. Due to the normalization of the wavefunction, we have $\sum_m \lvert \phi_m \rvert^2 = 1$, where
\begin{equation}
\sum_m \lvert \phi_m \rvert^2 = c\frac{1 - \lvert\xi\rvert^{2N}}{1 - \lvert\xi\rvert^2} \xrightarrow{N\to\infty} \frac{c}{1 - \lvert\xi\rvert^2} = 1,
\end{equation}
so that $c=1-\lvert\xi\rvert^2$. Therefore, in thermodynamic limit,
\begin{equation}
\mathrm{IPR}_\mathrm{exp} = c^2 (1 + \lvert\xi\rvert^4 + \dots + \lvert\xi\rvert^{4N-4}) = \frac{(1 - \lvert\xi\rvert^2) (1 - \lvert\xi\rvert^{4N})}{1 + \lvert\xi\rvert^2}\xrightarrow{N\to\infty} \frac{1 - \lvert\xi\rvert^2}{1 + \lvert\xi\rvert^2} .
\end{equation}

Second, we study the IPRs of the extended SSH chains and then calculate the standard IPRs for localized topological states in transformed aperiodic chains. For the extended SSH chain with large winding numbers shown in Fig.~\ref{Fig-1}A in the main text, the nearest-neighbor hoppings are $w_{-1}$ and $w_0$, and the third-nearest-neighbor hoppings are $w_{-2}$ and $w_1$. As in the main text, the hopping amplitudes are taken as $w_{-1}=w_0=w_1=a$ and $w_{-2}=b$; for convenience we define a parameter $s=a/b$, which satisfies $0<s<1$ when the winding number equals two. The extended SSH chain has a total of $2N$ sites; due to the chiral symmetry, for the edge state $\psi$ localized at the left boundary, the components of the wavefunction on the even-numbered sites are almost zero; then we let the component of the wavefunction on the odd-numbered site, i.e., the $(2m-1)$th site, be denoted as $\varphi_m$. Thus, we obtain
\begin{equation}\label{add1}
\begin{bmatrix}
	\varphi_{m+1}\\
	\varphi_{m+2}\\
	\varphi_{m+3}
\end{bmatrix}
=
\begin{bmatrix}
	0 & 1 & 0\\
	0 & 0 & 1\\
	-s & -s & -s
\end{bmatrix}
\begin{bmatrix}
	\varphi_{m}\\
	\varphi_{m+1}\\
	\varphi_{m+2}
\end{bmatrix}.
\end{equation}
For the wavefunction exponentially decaying from the left boundary, by letting $\varphi_{m+1} = \xi \varphi_{m}$ ($\lvert\xi\rvert <1$), Eq.~\eqref{add1} is rewritten into an eigenvalue equation:
\begin{equation}\label{add2}
\begin{bmatrix}
	0 & 1 & 0\\
	0 & 0 & 1\\
	-s & -s & -s
\end{bmatrix}
\begin{bmatrix}
	\varphi_{m}\\
	\varphi_{m+1}\\
	\varphi_{m+2}
\end{bmatrix}
= \xi
\begin{bmatrix}
	\varphi_{m}\\
	\varphi_{m+1}\\
	\varphi_{m+2}
\end{bmatrix}.
\end{equation}
Substituting the hopping amplitudes $a=12$, $b=33$ in the main text into Eq.~\eqref{add2}, we obtain $\xi_1=-0.6554$ and $\xi_{2,3}=0.1459 \pm 0.7305\mathrm{i}$. (The final wavefunction should in principle be written as a superposition of the three exponentially localized states with decay coefficients $\xi_1$, $\xi_2$ and $\xi_3$; however, for an approximate evaluation, we shall only consider each separate state when calculating the standard IPR.) Turning to the transformed aperiodic chains, if we allow localization to occur at a maximum number $n_\mathrm{l}$ of separate sites in each mode, then the standard IPRs for aperiodic chains can be defined as
\begin{equation}
\mathrm{IPR}_\mathrm{st}^{(j)} = \frac{1 - \lvert \xi_j \rvert^2}{n_\mathrm{l} (1 + \lvert \xi_j \rvert^2)} ,
\end{equation}
in accordance with those exponentially localized states of the extended SSH chains. For the states of the transformed local aperiodic chains, if their calculated IPRs are larger than the smallest of $\mathrm{IPR}_\mathrm{st}^{(j)}$, we claim that these topological states are localized. In the main text, for the extended SSH chains with hopping amplitudes $a=12$, $b=33$, the standard IPRs are calculated to be $\mathrm{IPR}_\mathrm{st}^{(1)} = 0.1995$, and $\mathrm{IPR}_\mathrm{st}^{(2,3)} = 0.1432$ (where we choose $n_\mathrm{l}=2$); comparing the IPRs of the topological states of the transformed local aperiodic chain ($\mathrm{IPR}=0.3268$ and $0.3707$) with the standard IPRs, we find they can indeed be considered localized.

\subsection{Implementation of improved Householder method for moderately large lattice}\label{z4}
\begin{figure}[tbp]
	\centering
	\includegraphics[width=0.9\linewidth]{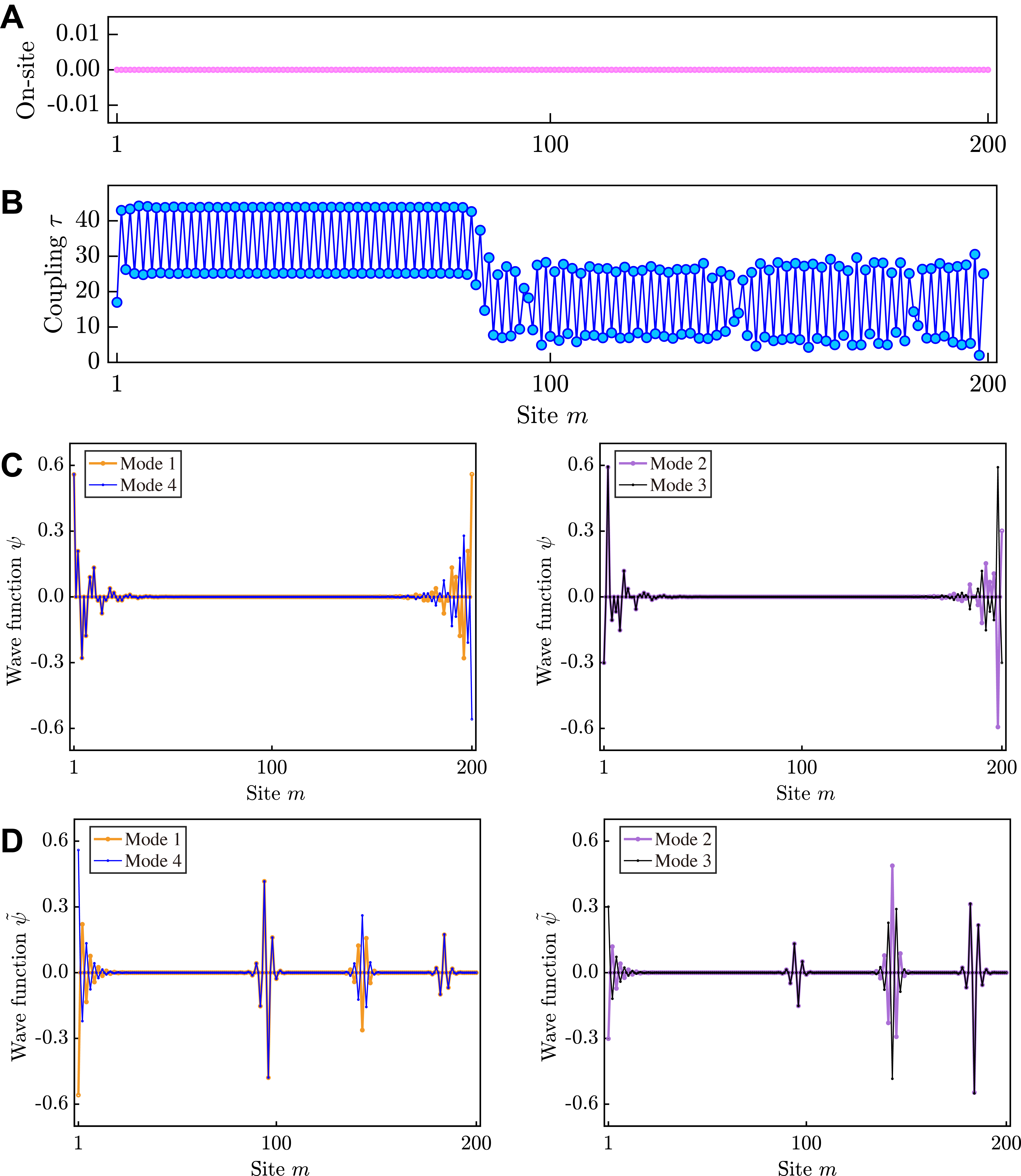}
	\caption{\label{Fig-sm5} \textbf{Numerical stability of improved Householder Method (100 unit cells in a chiral-symmetric chain).}  (\textbf{A}) On-site potentials of transformed aperiodic chain are all strictly zero. (\textbf{B}) Nearest couplings of transformed aperiodic chain. (\textbf{C}) Topological modes of Hamiltonian $H$ before transformation. (\textbf{D}) Topological modes of Hamiltonian $H_\mathrm{eff}$ after transformation. }
\end{figure}

The improved Householder method is generally the most stable tridiagonalization algorithm numerically; it is very effective for small artificial lattice structures, and is still accurate for systems with a moderately large number of unit cells. Aperiodic topological chains with artificial dimensions typically have about 20 sites (\textit{44, 45}); our improved Householder method is accurate for relatively large chains with more than 100 sites, and their geometric parameters are quite simple due to the zero on-site potentials, as proved below. We take the hopping parameters $w_{-2}=33, w_{-1/0/1}=12$ used in the main text for demonstration. The improved Householder method is applied on a 1D extended SSH chain consisting of 100 unit cells, whose Hamiltonian matrix has dimensions of $200\times200$. The on-site potentials of the transformed aperiodic chain are all strictly equal to zero as shown in Fig.~\ref{Fig-sm5}A, and the nearest-neighbor couplings are shown in Fig.~\ref{Fig-sm5}B, which is still easy to be fabricated according to Section Materials and Methods.
Compared with the precise result proved by Theorem~\ref{Theorem-2} that the on-site potentials for chiral-symmetric systems are zero, the example shows that the improved
Householder method is very stable on the computational side. Next, we examine the numerical stability for the results of the modes. Since the extended SSH chain before transformation has winding number 2, it has four localization sites at the edges, i.e., sites 1, 3, 198, and 200, as shown in Fig.~\ref{Fig-sm5}C. The structure has two pairs of topological modes, where the two modes are chiral-symmetric partners, that is, the absolute values of the components of the modes at each site are the same. After mapping the Hamiltonian $H$ of the extended SSH chain to the Hamiltonian $H_\mathrm{eff}$ of the aperiodic chain, as shown in Fig.~\ref{Fig-sm5}D, the topological modes become well localized at sites 1, 96, 143, and 184, and the two topological states in each pair are still chiral-symmetric partners themselves. There is almost no error in the eigenvalues before and after the transformation, and the eigenvectors are also unitarily transformed as Eq.~\eqref{eqb:3}, indicating that the method is numerically stable.

\subsection{Equivalent topological invariant after tridiagonalization}\label{e}
The topological property of the finite extended SSH chain (Hamiltonian $H$) can be characterized by LTM in Section~\ref{a}. After tridiagonalization, the bipartite extended SSH chain is mapped onto a new aperiodic chain with nearest-neighbor couplings  (Hamiltonian $H_{\mathrm{eff}}$). In virtue of the values of $V$ and $H_{\mathrm{eff}}$, the topological invariant of the mapped aperiodic chain is
\begin{equation}
\label{eqd:1}
\tilde{\gamma}(l)=\frac{1}{2} \sum\limits_{a=A,B} V\{(\tilde{Q}_{BA}[\tilde{X},\tilde{Q}_{AB}])_{la,la} + (\tilde{Q}_{AB}[\tilde{Q}_{BA},\tilde{X}])_{la,la} \}V^{\dagger}.
\end{equation}
We prove that $\tilde{\gamma}(l)$ in Eq.~\eqref{eqd:1} for the mapped aperiodic chains is equal to $\gamma(l)$ in Eq.~\eqref{eqa:5} for the finite extended SSH chains in the following. Firstly, the multiple zero modes of the mapped aperiodic chain are protected by the special nonlocal symmetry $\tilde{\Gamma}=V^{\dagger} \Gamma V$, which makes the tridiagonal Hamiltonian $\tilde{H}$ satisfy $\tilde{\Gamma}\tilde{H}=-\tilde{H}\tilde{\Gamma}$. Given $\Gamma=\Gamma_{A}-\Gamma_{B}$, we have $\tilde{\Gamma}_a=V^{\dagger} \Gamma_a V~(a\in\{A,B\})$. Secondly, considering $\tilde{\psi}=V^{\dagger}\psi$, it follows that $\tilde{P}_{\pm}=\tilde{\psi}_{\pm}\tilde{\psi}_{\pm}^{\dagger}=V^{\dagger}\psi_{\pm}(V^{\dagger}\psi_{\pm})^{\dagger}=V^{\dagger}(\psi_{\pm}\psi_{\pm}^{\dagger})V=V^{\dagger}P_{\pm}V$, so $\tilde{Q}=\tilde{P}_{+}-\tilde{P}_{-}=V^{\dagger}QV$. Finally, substituting $\tilde{Q}_{AB}=V^{\dagger}Q_{AB}V$, $\tilde{Q}_{BA}=V^{\dagger}Q_{BA}V$ and $\tilde{X}=V^{\dagger}XV$ into Eq.~\eqref{eqd:1}, we find $\tilde{\gamma}(l)=\gamma(l)$.

\subsection{Simulated topological modes by selected excitations}\label{h}
\begin{figure}[tbp]
	\centering
	\includegraphics[width=0.7\linewidth]{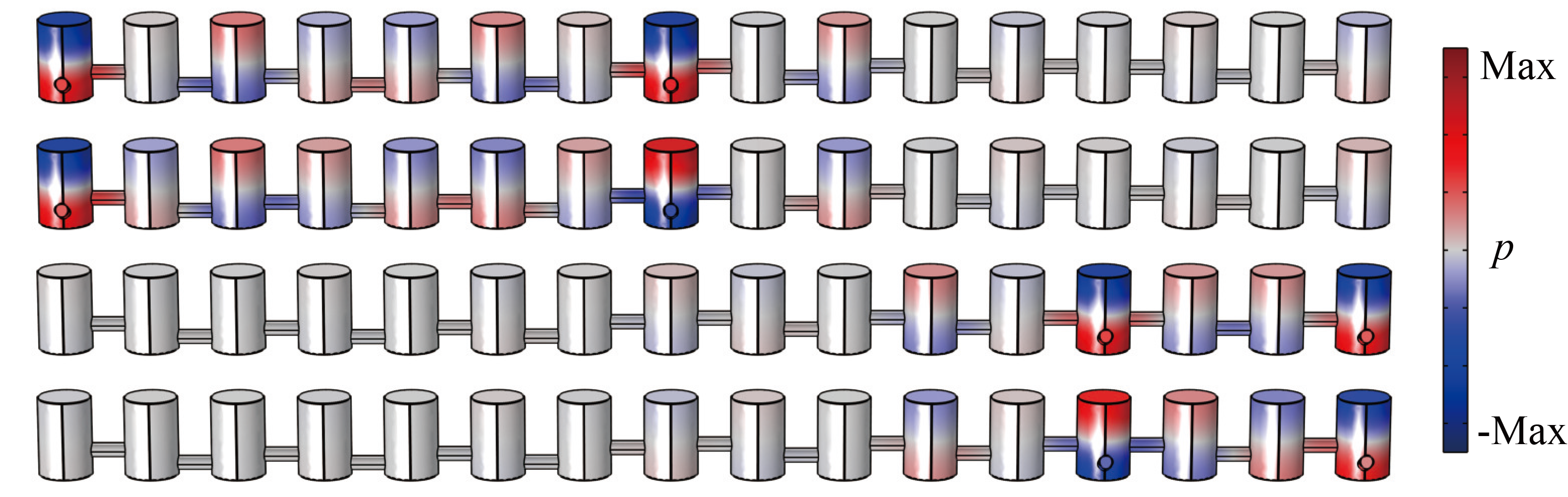}
	\caption{\label{Fig-sm2} \textbf{The four topological states achieved by selected excitations, calculated by finite-element software COMSOL Multiphysics.} Two of the modes are localized at the first and eighth sites but have different mode symmetries at these sites; the other two are localized at the thirteenth and sixteenth sites.}
\end{figure}

The excited topological states are simulated by the finite-element software COMSOL Multiphysics. As shown in Fig.~\ref{Fig-1}F, two of the four topological eigenmodes are localized at the first and eighth sites (the other two at the thirteenth and sixteenth sites), but have different mode symmetries (phase distributions) at these sites. Because the microphone can only measure the absolute sound pressure, we design a special method of excitation to distinguish the two states where one state has the same amplitude distribution but a different phase distribution from the other state. To this end, small ports are opened in the front of the first and eighth (thirteenth and sixteenth) sites as shown in Fig.~\ref{Fig-sm2}, and are excited by in-phase and anti-phase sources with equal amplitudes. The simulated results in Fig.~\ref{Fig-sm2} indicate that the four topological states can be achieved by selected excitations, which are consistent with the experimental results in Fig.~\ref{Fig-3}.

\subsection{Topological corner states of the 2D extended SSH model}\label{z3}
\begin{figure}[tbp]
	\centering
	\includegraphics[width=0.8\linewidth]{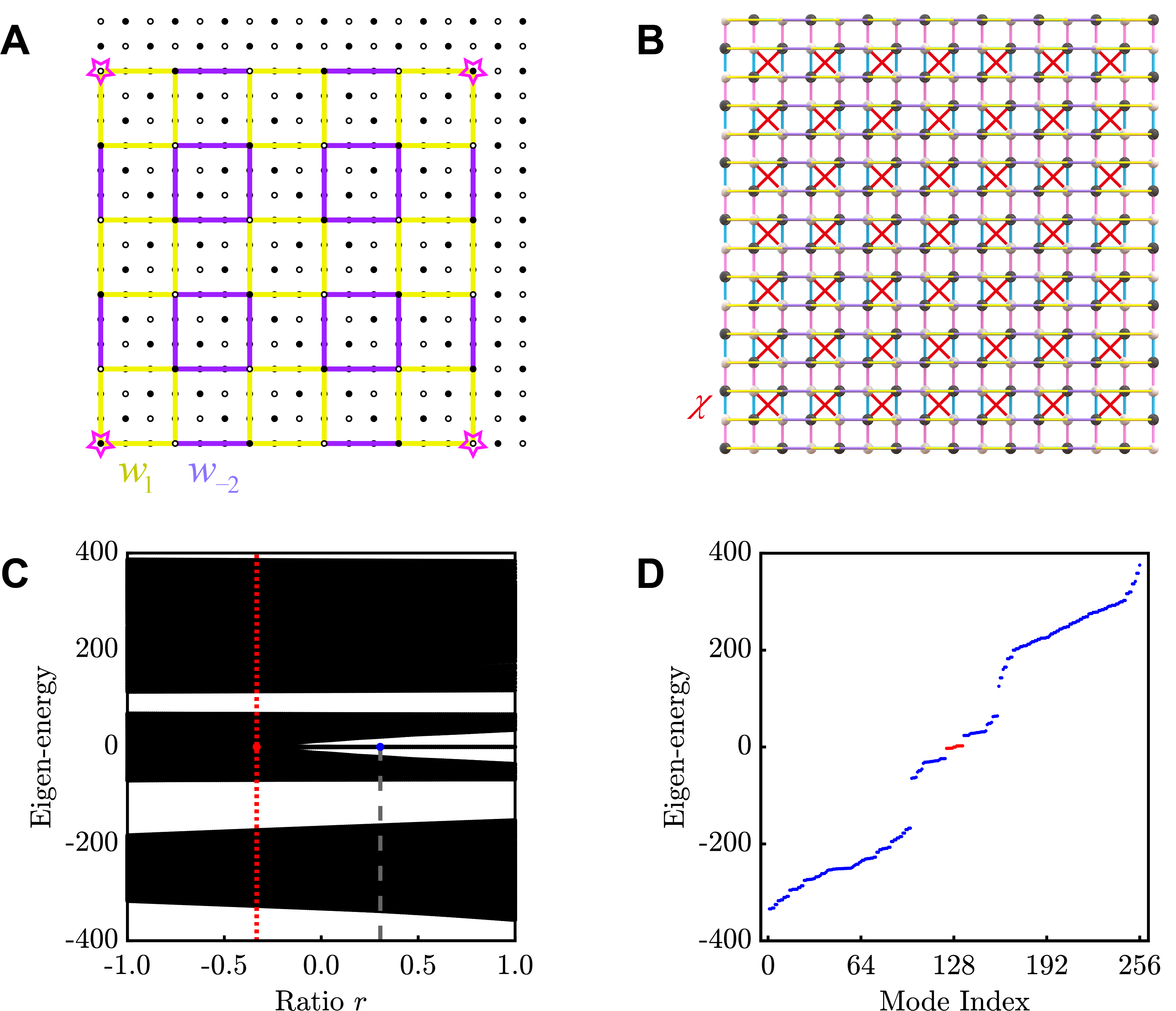}
	\caption{\label{Fig-6} \textbf{Analysis of the topological corner states of the 2D extended SSH model.} (\textbf{A}) The extreme case in which the nearest-neighbor hoppings of the 2D extended SSH model are 0. The pink stars represent the localization sites of the  corner states for the 2D SSH model. (\textbf{B}) The 2D extended SSH model [the top view of Fig.~\ref{Fig-4}(A)] adding the next-nearest-neighbor hoppings $\chi$ in red color. (\textbf{C}) Energy spectrum of the 2D extended SSH model with the next-nearest-neighbor hoppings $\chi=250$. The abscissa represents the relative magnitude of the third-nearest-neighbor hoppings and the nearest-neighbor hoppings. The red dashed line $r =- 1/3$ represents the topological phase transition, and the gray dashed line represents the situation considered in our paper, i.e., $w_{-1}=w_0=w_1=12, w_{-2}=33$. (\textbf{D}) Energy spectrum corresponding to the gray dashed line in (C), where twelve red points at zero energy represent the topological corner states.}
\end{figure}

The topological properties of the 2D extended SSH model are further analyzed in this section. We consider the extreme case of the 2D extended SSH model, where the nearest-neighbor hoppings are all 0 ($w_{-1}=w_0=0$), and only the long-range hoppings are retained (let $w_{-2}> w_1$). The extreme case corresponds to nine 2D SSH models (\textit{50}), four of which are topologically nontrivial. Each topologically nontrivial 2D SSH model, such as Fig.~\ref{Fig-6}A, has four topological corner states. Rotating the system in Fig.~\ref{Fig-6}A around the center point by $\pi/4$, $\pi/2$, and $3\pi/4$ yield another three nontrivial 2D SSH models, so the extreme case of the 2D extended SSH model has sixteen topological corner states in total.
Next, we consider the connection between the extreme case and the 2D extended SSH model with the parameters used in this work. By adding the next-nearest-neighbor hoppings $\chi$ shown in Fig.~\ref{Fig-6}B to the system in Fig.~\ref{Fig-4}A, the energy spectrum of the system in Fig.~\ref{Fig-4}A can be opened at zero energy to distinguish corner states from the bulk states. In this way, we can clearly see the topological phase transition. However, by adding the next-nearest-neighbor hoppings, only twelve corner states will be retained in the extreme case, and the localization sites are $(m,n)\in \{1,3,14,16\}\times \{1,3,14,16\}$, except the sites $(3,3)$, $(3,14)$, $(14,3)$ and $(14,14)$. When we keep $w_{-1}/w_0=1$ and $w_{-2}/w_1=33/12$ constant, and define $r=(w_{-2}+w_1- w_{-1}- w_0)/\sum\nolimits_{n=-2}^{1} w_n$, by adjusting the value of $r$, we can change the relative magnitude of the nearest-neighbor and the third-nearest-neighbor hopping; then the energy spectrum is obtained as shown in Fig.~\ref{Fig-6}C. We find that the topological phase transition occurs at $r=-1/3$, from $r=1$ (corresponding to the extreme case where the nearest-neighbor hoppings are 0) to $r =  7/23$ (corresponding to the parameters $w_{-1}=w_0=w_1=12, w_{-2}=33$ used in our work), the energy spectrum is adiabatically connected. When $r=7/23$, the energy spectrum of the model in Fig.~\ref{Fig-6}B is shown in Fig.~\ref{Fig-6}D, where twelve red points correspond to twelve topological corner states. Thus, by establishing a connection to the 2D SSH model, we once again illustrate the novel topological states of the 2D extended SSH model, together with those of the equivalent simple 2D aperiodic structure.

\subsection{Simulated eigenfrequencies and eigenmodes in 2D system}\label{z1}
\begin{figure}[tbp!]
	\centering
	\includegraphics[width=\linewidth]{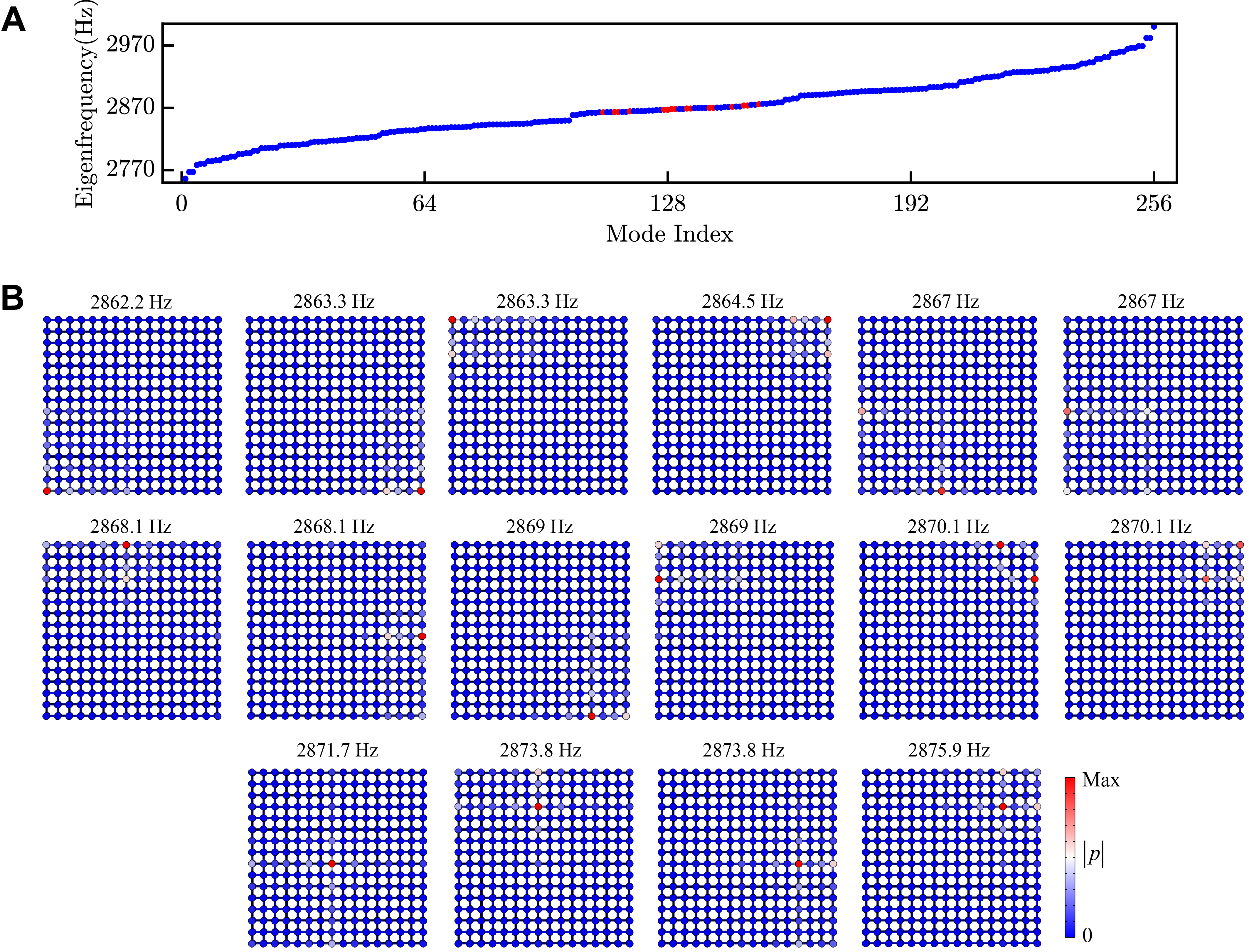}
	\caption{\label{Fig-sm3} \textbf{The eigenfrequencies and eigenmodes of the 2D practical structure.} (\textbf{A}) The frequency spectrum where red dots correspond to the eigenfrequencies of sixteen 0D topological states. (\textbf{B}) The sound pressure distributions of the sixteen 0D topological states.}
\end{figure}

We construct a 2D practical structure in Fig.~\ref{Fig-4}F according to the 2D local aperiodic toy model in Fig.~\ref{Fig-4}D. The eigenfrequencies and eigenmodes of the practical structure are given in Fig.~\ref{Fig-sm3} by COMSOL Multiphysics. The frequency spectrum of the practical structure in Fig.~\ref{Fig-sm3}A well emulates the energy spectrum of the 2D local aperiodic model in Fig.~\ref{Fig-4}C; the sixteen eigenmodes corresponding to the 0D topological states shown in Fig.~\ref{Fig-sm3}B localize at the sixteen sites as in Fig.~\ref{Fig-4}F.

\subsection{Simulated topological states in 2D system by selected excitations}\label{z2}
\begin{figure}[tbp!]
	\centering
	\includegraphics[width=0.8\linewidth]{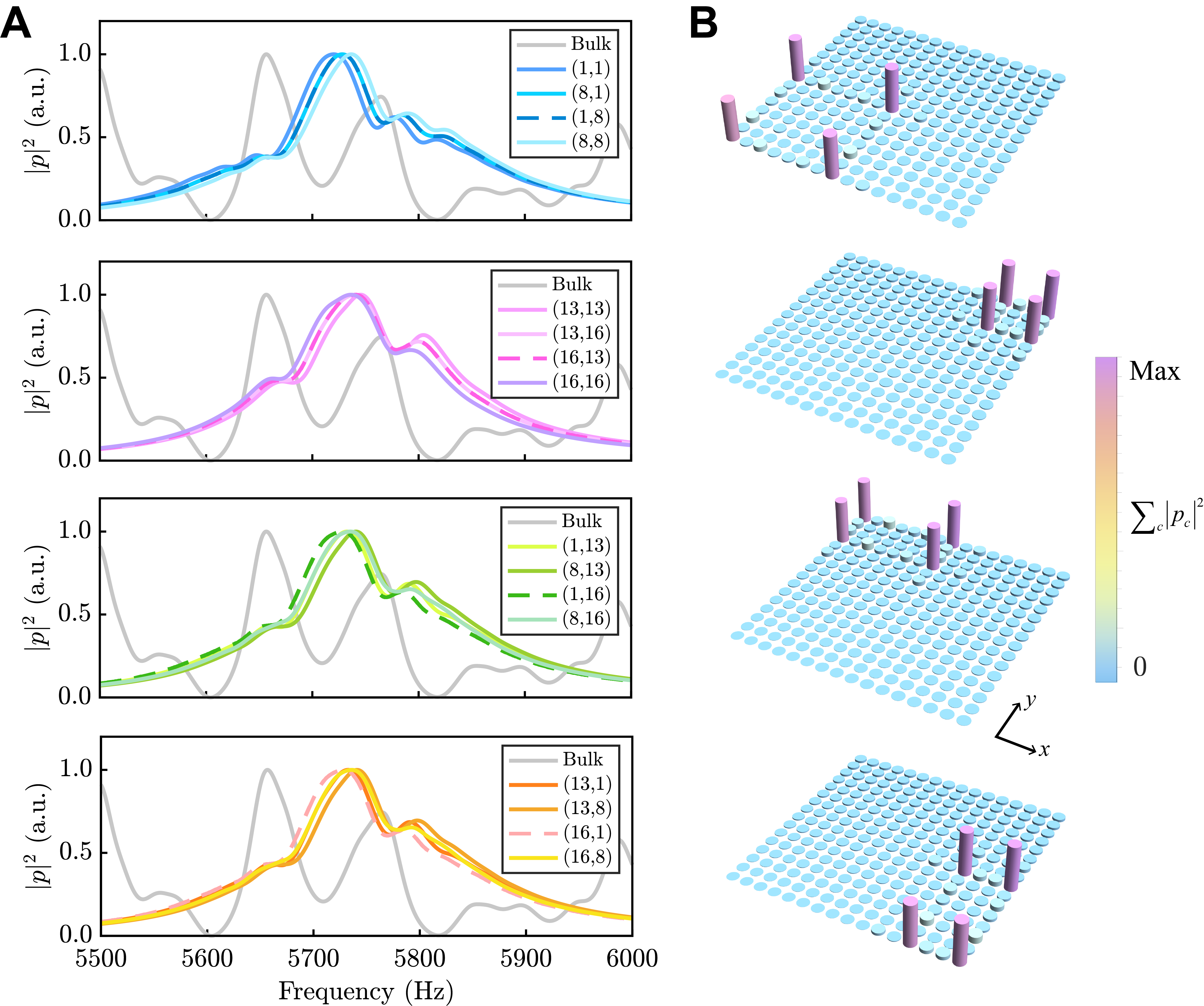}
	\caption{\label{Fig-sm4} \textbf{The simulated topological states of 2D aperiodic structure by selected excitations.} (\textbf{A}) The acoustic intensity response spectra of 0D topological states with excitations at sites $(m,n)\in\{1,8,13,16\}\times \{1,8,13,16\}$, where four sites are drawn in a group, and also a nonlocalized bulk mode for comparison. (\textbf{B}) The sound field distributions of the sixteen 0D topological states.}
\end{figure}

Considering thermoviscous loss by COMSOL Multiphysics, we simulate the 0D topological states under selected excitations, thereby comparing 
the results with the experimental ones. The practical structure of the 2D finite-element simulation is exactly the same as that of the 2D experiment. In the first step, we give excitation at the 16 designated sites [site $(m,n)\in\{1,8,13,16\}\times\{1,8,13,16\}$], and measure the sound pressure at the excitation sites to obtain the acoustic intensity response spectra. When a certain site is selected for excitation, the resonator corresponding to this site in Fig.~\ref{Fig-5}A is set as the resonator with the curved tube in Fig.~\ref{Fig-7}F, but the other resonators remain unchanged; and the port excitation condition is added at the upper end of the curved tube. Then, the sound pressure at the center of the top surface of this resonator in the range of 5500--6000 Hz is collected to obtain the simulated result as shown in Fig.~\ref{Fig-sm4}A.
It is seen that obvious peaks appear in the range of 5728--5745 Hz. At the same time, we give excitation at site $(4,4)$, but collect sound pressure at site $(3,4)$, so as to obtain the sound response spectrum of the nonlocalized bulk mode, and find that its peak is obviously staggered from the localized 0D topological state. In the second step, we again excite each of the 16 designated sites, but collect the sound pressure of all sites at the frequencies of the peaks in Fig.~\ref{Fig-sm4}A, and plot the simulated results as shown in Fig.~\ref{Fig-sm4}B. We find that the sound field distributions shown in Fig.~\ref{Fig-sm4}B are obviously localized at sixteen sites, and the 2D practical structure has sixteen 0D topological states.

\begin{thebibliography}{10}
		
		\bibitem{Hasan2010}
		M.~Z. Hasan, C.~L. Kane, Colloquium: Topological insulators. {\it Rev. Mod. Phys.\/} {\bf 82}, 3045 (2010).
		
		\bibitem{ZhangSC2011}
		X.-L. Qi, S.-C. Zhang, Topological insulators and superconductors. {\it Rev. Mod. Phys.\/} {\bf 83}, 1057 (2011).
		
		\bibitem{semimetal2018}
		N.~P. Armitage, E.~J. Mele, A.~Vishwanath, Weyl and Dirac semimetals in three-dimensional solids. {\it Rev. Mod. Phys.\/} {\bf 90},
		015001 (2018).
		
		\bibitem{Schnyder2008}
		A.~P. Schnyder, S.~Ryu, A.~Furusaki, A.~W.~W. Ludwig, Classification of topological insulators and superconductors in three spatial dimensions. {\it Phys. Rev. B\/} {\bf
			78}, 195125 (2008).
		
		\bibitem{Chiu2016}
		C.-K. Chiu, J.~C.~Y. Teo, A.~P. Schnyder, S.~Ryu, Classification of topological quantum matter with symmetries. {\it Rev. Mod. Phys.\/} {\bf
			88}, 035005 (2016).
		
		\bibitem{sanjuti-jixian}
		M.~Ezawa, Higher-Order topological insulators and semimetals on the breathing kagome and pyrochlore lattices. {\it Phys. Rev. Lett.\/} {\bf 120}, 026801 (2018).
		
		\bibitem{photonic2008}
		F.~D.~M. Haldane, S.~Raghu, Possible realization of directional optical waveguides in photonic crystals with broken time-reversal symmetry. {\it Phys. Rev. Lett.\/} {\bf 100}, 013904 (2008).
		
		\bibitem{photonic2013}
		M.~C. Rechtsman, {\it et~al.\/}, Photonic Floquet topological insulators. {\it Nature\/} {\bf 496}, 196 (2013).
		
		\bibitem{Photonic2021}
		Z.-Q. Jiao, {\it et~al.\/}, Experimentally detecting quantized Zak phases without chiral symmetry in photonic lattices. {\it Phys. Rev. Lett.\/} {\bf 127}, 147401 (2021).
		
		\bibitem{Wang2023}
		Z.~Wang, {\it et~al.\/}, Sub-symmetry-protected topological states. {\it Nat. Phys.\/} {\bf 19}, 992 (2023).
		
		\bibitem{newcornerstate}
		M.~Li, {\it et~al.\/}, Higher-order topological states in photonic kagome crystals with long-range interactions. {\it Nat. Photonics\/} {\bf 14}, 89 (2020).
		
		\bibitem{kirsch_nonlinear_2021}
		M.~S. Kirsch, {\it et~al.\/}, Nonlinear second-order photonic topological insulators. {\it Nat. Phys.\/} {\bf 17}, 995 (2021).
		
		\bibitem{pan_real_2023}
		Y.~Pan, {\it et~al.\/}, Real higher-order {Weyl} photonic crystal. {\it Nat. Commun.\/} {\bf 14}, 6636 (2023).
		
		\bibitem{Ma2019}
		G.~Ma, M.~Xiao, C.~T. Chan, Topological phases in acoustic and mechanical
		systems. {\it Nat. Rev. Phys.\/} {\bf 1}, 281 (2019).
		
		\bibitem{Xue2022}
		H.~Xue, Y.~Yang, B.~Zhang, Topological acoustics. {\it Nat. Rev. Mater.\/} {\bf 7}, 974 (2022).
		
		\bibitem{Xiao2015exp}
		M.~Xiao, {\it et~al.\/}, Geometric phase and band inversion in periodic acoustic
		systems. {\it Nat. Phys.\/} {\bf 11}, 240 (2015).
		
		\bibitem{Wang_2022}
		W.~Wang, X.~Wang, G.~Ma, Non-{H}ermitian morphing of topological modes. {\it Nature\/} {\bf 608}, 50 (2022).
		
		\bibitem{local-marker}
		X.~Shi, {\it et~al.\/}, Disorder-induced topological phase transition in a
		one-dimensional mechanical system. {\it Phys. Rev. Res.\/} {\bf 3}, 033012 (2021).
		
		\bibitem{highorder}
		H.~Chen, {\it et~al.\/}, Creating synthetic spaces for higher-order topological
		sound transport. {\it Nat. Commun.\/} {\bf 12}, 5028 (2021).
		
		\bibitem{PRLpumping}
		M.~I.~N. Rosa, R.~K. Pal, J.~R.~F. Arruda, M.~Ruzzene, Edge states and
		topological pumping in spatially modulated elastic lattices. {\it Phys. Rev. Lett.\/}
		{\bf 123}, 034301 (2019).
		
		\bibitem{nj-zs}
		Y.-F. Chen, {\it et~al.\/}, Various topological phases and their abnormal
		effects of topological acoustic metamaterials. {\it Interdiscip. Mater.\/} {\bf 2}, 179 (2023).
		
		\bibitem{pnas-mecha}
		M.~Guzman, X.~Guo, C.~Coulais, D.~Carpentier, D.~Bartolo, Model-free
		characterization of topological edge and corner states in mechanical
		networks. {\it Proc. Natl.
			Acad. Sci.\/} {\bf 121}, e2305287121 (2024).
		
		\bibitem{pumpelastic}
		S.~Wang, {\it et~al.\/}, Smart patterning for topological pumping of elastic
		surface waves. {\it Sci. Adv.\/} {\bf 9}, eadh4310 (2023).
		
		\bibitem{Jiang22}
		J.~Jiang, {\it et~al.\/}, Multiband topological states in non-{H}ermitian
		photonic crystals. {\it Opt. Lett.\/} {\bf 47}, 437 (2022).
		
		\bibitem{multi}
		K.-H. Kim, K.-K. Om, Multiband photonic topological valley-{H}all edge modes
		and second-order corner states in square lattices. {\it Adv. Opt. Mater.\/} {\bf 9}, 2001865 (2021).
		
		\bibitem{multipolar_lasing}
		H.-R. Kim, {\it et~al.\/}, Multipolar lasing modes from topological corner
		states. {\it Nat. Commun.\/} {\bf 11}, 5758 (2020).
		
		\bibitem{duotiao}
		Y.~Meng, {\it et~al.\/}, Spinful topological phases in acoustic crystals with
		projective {$PT$} symmetry. {\it Phys. Rev. Lett.\/} {\bf 130}, 026101 (2023).
		
		\bibitem{degenerate-disclination}
		Y.~Deng, {\it et~al.\/}, Observation of degenerate zero-energy topological
		states at disclinations in an acoustic lattice. {\it Phys. Rev. Lett.\/} {\bf 128}, 174301 (2022).
		
		\bibitem{Zhang2023}
		Q.~Zhang, {\it et~al.\/}, Observation of acoustic non-{H}ermitian Bloch braids
		and associated topological phase transitions. {\it Phys. Rev. Lett.\/} {\bf 130}, 017201 (2023).
		
		\bibitem{Z-the}
		W.~A. Benalcazar, A.~Cerjan, Chiral-symmetric higher-order topological phases
		of matter. {\it Phys. Rev. Lett.\/} {\bf 128}, 127601 (2022).
		
		\bibitem{Z-exp}
		D.~Wang, {\it et~al.\/}, Realization of a $\mathbb{Z}$-classified
		chiral-symmetric higher-order topological insulator in a coupling-inverted
		acoustic crystal. {\it Phys. Rev. Lett.\/} {\bf 131}, 157201 (2023).
		
		\bibitem{Floquetsum}
		R.~W. Bomantara, L.~Zhou, J.~Pan, J.~Gong, Coupled-wire construction of static
		and {F}loquet second-order topological insulators. {\it Phys. Rev. B\/} {\bf 99},
		045441 (2019).
		
		\bibitem{any-roton}
		A.~Kazemi, {\it et~al.\/}, Drawing dispersion curves: Band structure
		customization via nonlocal phononic crystals. {\it Phys. Rev. Lett.\/} {\bf 131}, 176101 (2023).
		
		\bibitem{indexing}
		R.~G. Dias, A.~M. Marques, Long-range hopping and indexing assumption in
		one-dimensional topological insulators. {\it Phys. Rev. B\/} {\bf 105}, 035102 (2022).
		
		\bibitem{Haldane2014}
		L.~Li, Z.~Xu, S.~Chen, Topological phases of generalized
		{S}u-{S}chrieffer-{H}eeger models. {\it Phys. Rev. B\/} {\bf 89}, 085111 (2014).
		
		\bibitem{chen_roton-like_2021}
		Y.~Chen, M.~Kadic, M.~Wegener, Roton-like acoustical dispersion relations in
		{3D} metamaterials. {\it Nat. Commun.\/} {\bf 12}, 3278 (2021).
		
		\bibitem{exp-roton-like}
		J.~A.~I. Martínez, {\it et~al.\/}, Experimental observation of roton-like
		dispersion relations in metamaterials. {\it Sci. Adv.\/} {\bf 7}, eabm2189 (2021).
		
		\bibitem{CM-nonlocal-2022}
		K.~Wang, Y.~Chen, M.~Kadic, C.~Wang, M.~Wegener, Nonlocal interaction
		engineering of {2D} roton-like dispersion relations in acoustic and
		mechanical metamaterials. {\it Commun. Mater.\/} {\bf
			3}, 35 (2022).
		
		\bibitem{CABS-prl}
		Z.-D. Zhang, M.-H. Lu, Y.-F. Chen, Observation of free-boundary-induced chiral
		anomaly bulk states in elastic twisted kagome metamaterials. {\it Phys. Rev. Lett.\/} {\bf 132}, 086302
		(2024).
		
		\bibitem{CHEN2020}
		B.-H. Chen, D.-W. Chiou, An elementary rigorous proof of bulk-boundary
		correspondence in the generalized {S}u-{S}chrieffer-{H}eeger model. {\it Phys. Lett. A\/} {\bf 384}, 126168 (2020).
		
		\bibitem{Electronlong2019}
		B.~P\'erez-Gonz\'alez, M.~Bello, A.~G\'omez-Le\'on, G.~Platero, Interplay
		between long-range hopping and disorder in topological systems. {\it Phys. Rev.
			B\/} {\bf 99}, 035146 (2019).
		
		\bibitem{Asbth2016}
		J.~K. Asb{\'{o}}th, L.~Oroszl{\'{a}}ny, A.~P{\'{a}}lyi, {\it A Short Course on
			Topological Insulators\/} (Springer, Cham, 2016).
		
		\bibitem{Maffei2018}
		M.~Maffei, A.~Dauphin, F.~Cardano, M.~Lewenstein, P.~Massignan, Topological
		characterization of chiral models through their long time dynamics. {\it New J.
			Phys.\/} {\bf 20}, 013023 (2018).
		
		\bibitem{lanczosnature}
		L.~J. Maczewsky, {\it et~al.\/}, Synthesizing multi-dimensional excitation
		dynamics and localization transition in one-dimensional lattices. {\it Nat. Photonics\/} {\bf 14}, 76 (2020).
		
		\bibitem{pnas2021}
		K.~Chen, {\it et~al.\/}, Nonlocal topological insulators: Deterministic
		aperiodic arrays supporting localized topological states protected by
		nonlocal symmetries. {\it Proc. Natl. Acad. Sci.\/} {\bf 118}, e2100691118
		(2021).
		
		\bibitem{householder}
		M.~H. Teimourpour, L.~Ge, D.~N. Christodoulides, R.~El-Ganainy, Non-{Hermitian}
		engineering of single mode two dimensional laser arrays. {\it Sci.
			Rep.\/} {\bf 6}, 33253 (2016).
		
		\bibitem{recipe}
		W.~H. Press, S.~A. Teukolsky, W.~T. Vetterling, B.~P. Flannery, {\it Numerical
			Recipes in C: The Art of Scientific Computing, Second Edition\/} (Cambridge
		University Press, Cambridge, 1992).
		
		\bibitem{zhu_time-periodic_2022}
		W.~Zhu, H.~Xue, J.~Gong, Y.~Chong, B.~Zhang, Time-periodic corner states from
		{Floquet} higher-order topology. {\it Nat. Commun.\/} {\bf 13}, 11
		(2022).
		
		\bibitem{Miert-zak-symmetry}
		G.~van Miert, C.~Ortix, C.~M. Smith, Topological origin of edge states in
		two-dimensional inversion-symmetric insulators and semimetals. {\it 2D Mater.\/} {\bf 4}, 015023 (2016).
		
		\bibitem{PRL2DSSH}
		F.~Liu, K.~Wakabayashi, Novel topological phase with a zero Berry curvature. {\it Phys. Rev. Lett.\/} {\bf 118}, 076803 (2017).
		
		\bibitem{science-polarization}
		W.~A. Benalcazar, B.~A. Bernevig, T.~L. Hughes, Quantized electric multipole
		insulators. {\it Science\/} {\bf 357}, 61
		(2017).
		
		\bibitem{PRBnnc-circuit}
		N.~A. Olekhno, {\it et~al.\/}, Experimental realization of topological corner
		states in long-range-coupled electrical circuits. {\it Phys. Rev. B\/} {\bf 105}, L081107 (2022).
		
		\bibitem{nnc-phasediagram}
		M.-S. Wei, {\it et~al.\/}, Topological laser with higher-order corner states in
		the 2-dimensional Su-Schrieffer-Heeger model. {\it Opt. Express\/} {\bf 31}, 3427 (2023).
		
		\bibitem{emission}
		M.~Landi, J.~Zhao, W.~E. Prather, Y.~Wu, L.~Zhang, Acoustic Purcell effect for
		enhanced emission. {\it Phys. Rev. Lett.\/}
		{\bf 120}, 114301 (2018).
		
		\bibitem{Mostafa1998}
		M.~Fatemi, J.~F. Greenleaf, Ultrasound-stimulated vibro-acoustic spectrography. {\it Science\/} {\bf 280}, 82 (1998).
		
		\bibitem{Epifanio_1999}
		C.~L. Epifanio, J.~R. Potter, G.~B. Deane, M.~L. Readhead, M.~J. Buckingham, Imaging in the ocean with ambient noise: The {ORB} experiments.
		{\it J. Acoust. Soc. Am.\/} {\bf 106}, 3211 (1999).
		
		\bibitem{computational-imaging}
		Y.~Altmann, {\it et~al.\/}, Quantum-inspired computational imaging. {\it Science\/} {\bf 361}, eaat2298 (2018).
		
		\bibitem{ZHANG201976}
		Q.~Zhang, Y.~Chen, K.~Zhang, G.~Hu, Programmable elastic valley Hall insulator
		with tunable interface propagation routes. {\it Extreme Mech. Lett.\/} {\bf 28}, 76
		(2019).
		
		\bibitem{zhang_low-threshold_2020}
		W.~Zhang, {\it et~al.\/}, Low-threshold topological nanolasers based on the
		second-order corner state. {\it Light: Sci. \& Appl.\/} {\bf 9}, 109 (2020).
		
		\bibitem{advs.201900401}
		H.~Liu, Q.~Zhang, K.~Zhang, G.~Hu, H.~Duan, Designing 3D digital metamaterial
		for elastic waves: From elastic wave polarizer to vibration control. {\it Adv. Sci.\/} {\bf 6}, 1900401
		(2019).
		
		\bibitem{unique}
		B.~N. Parlett, {\it The Symmetric Eigenvalue Problem\/} (Society for Industrial
		and Applied Mathematics, Philadelphia, 1998).
		
		\bibitem{localizationdegree}
		X.~Li, X.~Li, S.~Das~Sarma, Mobility edges in one-dimensional bichromatic
		incommensurate potentials. {\it Phys. Rev. B\/} {\bf 96}, 085119 (2017).
		
	\end{thebibliography}

\clearpage

\clearpage
\textbf{Movie S1.} Illustration movie of a programmable multimode topological saser under ``top-down'' excitations. By applying external sinusoidal excitations through small openings below the selected air cavities (which are at multiple sites: in the bulk, on the edge, or at the corner), the structure can realize \emph{sound amplification} at different sites, manifesting localized sound intensities at the selected cavities. The frequency of the external excitation $f = 1/T$ is approximately equal to the natural frequencies of the topological states. The black arrows represent the applied excitations at the present time.\\

\textbf{Movie S2.} Illustration movie of a programmable multimode topological saser under ``blockwise'' excitations. At this time the excitations are moved from one block to another in a continuous time frame. The damping of the air is adjusted to a larger value for better visual effects, demonstrating sound attenuation at those sites after removal of excitations more clearly.\\

\textbf{Movie S3.} Illustration movie of a programmable multimode topological saser under ``pointwise'' excitations.
	
\end{document}